# The Essential Role of Pair Matching in Cluster-Randomized Experiments, with Application to the Mexican Universal Health Insurance Evaluation

**Kosuke Imai, Gary King and Clayton Nall**

*Abstract.* A basic feature of many field experiments is that investigators are only able to randomize clusters of individuals—such as households, communities, firms, medical practices, schools or classrooms—even when the individual is the unit of interest. To recoup the resulting efficiency loss, some studies pair similar clusters and randomize treatment within pairs. However, many other studies avoid pairing, in part because of claims in the literature, echoed by clinical trials standards organizations, that this *matched-pair, cluster-randomization* design has serious problems. We argue that all such claims are unfounded. We also prove that the estimator recommended for this design in the literature is unbiased only in situations when matching is unnecessary; its standard error is also invalid. To overcome this problem without modeling assumptions, we develop a simple design-based estimator with much improved statistical properties. We also propose a model-based approach that includes some of the benefits of our design-based estimator as well as the estimator in the literature. Our methods also address individual-level noncompliance, which is common in applications but not allowed for in most existing methods. We show that from the perspective of bias, efficiency, power, robustness or research costs, and in large or small samples, pairing should be used in cluster-randomized experiments whenever feasible; failing to do so is equivalent to discarding a considerable fraction of one's data. We develop these techniques in the context of a randomized evaluation we are conducting of the Mexican Universal Health Insurance Program.

*Kosuke Imai is Assistant Professor, Department of Politics, Princeton University, Princeton, New Jersey 08544, USA e-mail: kimai@princeton.edu; URL: http://imai.princeton.edu. Gary King is Albert J. Weatherhead III University Professor, Institute for Quantitative Social Science, Harvard University, 1737 Cambridge St., Cambridge, Massachusetts 02138, USA e-mail: King@Harvard.edu; URL:*

*http://GKing.Harvard.edu. Clayton Nall is Ph.D. Candidate, Department of Government, Institute for Quantitative Social Science, Harvard University, 1737 Cambridge St., Cambridge, Massachusetts 02138, USA e-mail: nall@fas.harvard.edu.*

## 1. INTRODUCTION

For political, ethical or administrative reasons, researchers conducting field experiments are often unable to randomize treatment assignment to individuals and so instead randomize treatments to clusters of individuals (Murray, 1998; Donner and Klar, 2000a; Raudenbush, Martinez and Spybrook, 2007). For example, 19 (68%) of the 28 field experiments we found published in major political science journals since 2000 randomized households, precincts, city-blocks or villages even though individual voters were the inferential target (e.g., Arceneaux, 2005); in public health and medicine, where "the number of trials reporting a cluster design has risen exponentially since 1997" (Campbell, 2004), randomization occurs at the level of health clinics, physicians or other administrative and geographical units even though individuals are the units of interest (e.g., Sommer et al., 1986; Varnell et al., 2004); and numerous education researchers randomize schools, classrooms or teachers instead of students (e.g., Angrist and Lavy, 2002).

Since efficiency drops when randomizing clusters of individuals instead of individuals themselves (Cornfield, 1978), many scholars attempt to recoup some of this lost efficiency by pairing clusters, based on the similarity of available background characteristics, before randomly assigning one cluster within each pair to receive the treatment assignment (e.g., Ball and Bogatz, 1972; Gail et al., 1992; Hill, Rubin and Thomas, 1999). Since matching prior to random treatment assignment can greatly improve the efficiency of causal effect estimation (Bloom, 1978; Greevy et al., 2004), and matching in pairs can be substantially more efficient than matching in larger blocks, *matched-pair, cluster-randomization* (MPCR) would appear to be an attractive design for field experiments (Imai, King and Stuart, 2008). [See also Moulton (2004).] The design is especially useful for public policy experiments since, when used properly, it can be robust to interventions by politicians and others that have ruined many policy evaluations, such as when office-holders arrange program benefits for constituents who live in control group clusters (King et al., 2007).

Unfortunately, despite its apparent benefits and common usage, this experimental design has an uncertain scientific status. Researchers in this area and formal statements from clinical trial standards organizations (e.g., Donner and Klar, 2004; Feng et al., 2001; Medical Research Council, 2002) claim that certain "analytic limitations" make MPCR, or at least the existing methods available to analyze data from this design, inappropriate. These claimed limitations include "the restriction of prediction models to cluster-level baseline risk factors (e.g., cluster size), the inability to test for homogeneity of . . . [causal effects across clusters], and difficulties in estimating the intracluster correlation coefficient, a measure of similarity among cluster members" (Klar and Donner, 1997, page 1754). In addition, in a widely cited article, Martin et al. (1993) claim that in small samples, pairing can reduce statistical power.

We show that each of the claims regarding analytical limitations of MPCR is incorrect. We also demonstrate that the power calculations leading Martin et al. (1993) to recommend against MPCR in small samples is dependent on an assumption of equal cluster sizes that vitiates one major advantage of pair matching; we show in real data that the assumption does not apply and without it pair matching on cluster sizes and pre-treatment variables that affect the outcome improves both efficiency and power a great deal, even in samples as small as three pairs. In fact, because the efficiency gain of MPCR depends on the correlation of cluster means weighted by cluster size, the advantage can be much larger than the unweighted correlations that have been studied seem to indicate, even when cluster sizes are independent of the outcome.

Finally, there exists no published formal evaluation of the statistical properties of the estimator for MPCR data most commonly recommended in the methodological literature. By defining the quantities of interest separately from the methods used to estimate them, and identifying a model that gives rise to the most commonly used estimator, we show that this approach depends on assumptions, such as the homogeneity of treatment effects across all clusters, that apply best when matching is not needed to begin with. The commonly used variance estimator is also biased. We then offer new simple design-based estimators and their variances. We also propose an alternative model-based approach that includes the benefits of our design-based estimator, which has little or no bias, and the estimator in the literature, which under certain circumstances has lower variance. Finally, we extend our methods to situations with individual-level noncompliance, which is a basic feature of many MPCR experiments but for which most prior methods have not been adapted. With the results and new methods offered here, ambiguity about what to do in cluster randomized experiments vanishes: *pair matching should be used whenever feasible.*

## 2. EVALUATION OF THE MEXICAN UNIVERSAL HEALTH INSURANCE PROGRAM

As a running example of MPCR, we introduce a randomized evaluation we are conducting of *Seguro Popular de Salud* (SPS) in Mexico. A major domestic initiative of the Vicente Fox presidency, the program seeks "to provide social protection in health to the 50 million uninsured Mexicans" (Frenk et al., 2003, page 1667), constituting about half the population (King et al., 2007). The government intends to spend an additional one percent of GDP on health compared to 2002 once the program is fully introduced.

SPS permitted a cluster randomized (CR) study to be built into the program rollout. Under national legislation, Mexican states must apply to the federal government for funds both to publicize the program and fund its operations. The federal government approves these requests only when local health clinics are brought up to federal standards. When an area is approved to begin program enrollment, families who affiliate are expected to receive free preventative and regular medical care, pharmaceuticals and medical procedures. However, because local health clinics and hospitals may take years to meet federal standards, and also because of budget restrictions, a staged rollout was necessary and also allowed us the chance to run this randomized study. Finally, since SPS allows individuals to decide for themselves whether to enroll (if necessary, by traveling from unenrolled to enrolled areas), it was possible to adopt a clustered encouragement design (Frangakis et al., 2002), thereby permitting estimation of individual-level program effects. (We focus on the ITT effect until Section 6.)

The MPCR design was implemented in geographic areas created for the project which we call "health clusters," defined as the geographic catchment area of a local hospital or clinic. The country is tiled by 12,824 such clusters, and negotiations with the Mexican government produced more than 100 for which random assignment was acceptable. The chosen clusters were paired based on census demographics, poverty, education, and health infrastructure. Within each pair, one "treatment" cluster was randomly chosen for early program rollout, receiving funds to upgrade their health clinics and encourage individual enrollment. The "control" cluster in each pair had its rollout set for some future time. (Individuals could still obtain SPS benefits by traveling to SPS-approved clusters, but did not receive encouragement or resources to do so.) For design details, see King et al. (2007).

The primary outcome of interest at this stage was the level of out-of-pocket health expenditures, while secondary outcomes of interest included medical utilization, health self-assessment and self-reported health behaviors. Outcomes were measured in a baseline and followup panel survey of more than 32,000 households. Our examples draw upon 67 of these variables measured at the 10-month followup.

## 3. MATCHED-PAIR, CLUSTER-RANDOMIZED EXPERIMENTS

We now introduce MPCR experiments, including the theories of inference commonly applied (Section 3.1), the formal definitions, notation and assumptions used in (Section 3.2), and the quantities of interest typically sought (Section 3.3).

### 3.1 Theories of Inference

We describe the model-based and permutation-based theories of statistical inference that have been applied to MPCR data and then the design-based theory from which our work is derived.

First, model-based inference applied to MPCR typically uses generalized mixed-effects models, generalized estimating equations or multi-level models (Feng et al., 2001). Most of these work only if the modeling assumptions are correct; they also rely on asymptotic approximations. Model-based and model-assisted approaches have proved to be powerful in other areas, especially in survey research and missing data where it is often necessary, but they violate the purpose and spirit of experimental work which goes to great lengths and expense to avoid these types of assumptions.

Fisher's (1935) permutation-based theory of inference, which constructs exact nonparametric hypothesis tests based only on the random treatment assignment, has also been applied to MPCR. Although permutation inference in principle requires no models or approximations, in practice applications typically have required additional assumptions such as constant treatment effects across clusters or some kind of (e.g., Monte Carlo, large sample) approximations. The existing applications include Gail et al. (1996) and Braun and Feng (2001), which combine permutation inference with parametric modeling, and Small, Ten Have and Rosenbaum (2008) which considers quantile effects using different and more modest assumptions.

In contrast, we use Neyman's (1923) theory of inference, which is well known but has not before been attempted for MPCR. Like Fisher's permutation-based theory, Neyman's approach is also design (or "randomization") based and nonparametric, but it naturally avoids the constant treatment effect assumption and can provide valid inferences about both sample and population average treatment effects without modeling assumptions (Rubin, 1991). The estimators we derive are also simple to understand and easier to compute (requiring only weighted means and no numerical optimization, or simulation).

### 3.2 Formal Design Definition, Notation and Assumptions

Consider a MPCR experiment where $2m$ clusters are paired, based on a known function of the cluster characteristics, prior to the randomization of a binary treatment. We assume the $j$th cluster in the $k$th pair contains $n_{jk}$ units, where $j = 1, 2$ and $k = 1, \ldots, m$, and thus the total number of units is equal to $n = \sum_{k=1}^{m}(n_{1k} + n_{2k})$.

Under MPCR, simple randomization of an indicator variable, $Z_k$ for $k = 1, 2, \ldots, m$, is conducted independently across the $m$ pairs. For a pair with $Z_k = 1$, the first cluster within pair $k$ is treated (in our case, assigned encouragement to affiliate with SPS), and the second cluster is assigned control. In contrast, for a pair with $Z_k = 0$, the first cluster is the control whereas the second is treated. Thus, using $T_{jk}$ for the treatment indicator for the $j$th cluster in the $k$th pair, then $T_{1k} = Z_k$ and $T_{2k} = 1 - Z_k$. In the context of the SPS evaluation, we consider an intention-to-treat (ITT) analysis to estimate the causal effects of encouragement to affiliate with the program (see Section 6 on the estimation of causal effects of the actual affiliation).

We denote $Y_{ijk}(T_{jk})$ as the potential outcomes under the treatment ($T_{jk} = 1$) and control ($T_{jk} = 0$) conditions for the $i$th unit in the $j$th cluster of the $k$th pair (Holland, 1986; Maldonado and Greenland, 2002). The observed outcome variable is $Y_{ijk} = T_{jk}Y_{ijk}(1) + (1 - T_{jk})Y_{ijk}(0)$. Finally, the order of clusters within each pair is randomized so that the population distribution of $(Y_{i1k}(1), Y_{i1k}(0))$ equals $(Y_{i2k}(1), Y_{i2k}(0))$ (though this equality may not hold in sample).

A defining feature of CR experiments is that the potential outcomes for the $i$th unit in the $j$th cluster of the $k$th pair are a function of the cluster-level randomized treatment variable, $T_{jk}$, rather than its unit-level treatment counterpart. Similarly, the unit-level causal effect, $Y_{ijk}(1) - Y_{ijk}(0)$, is the difference between two unit-level potential outcomes that are the functions of the cluster-level treatment variable. Thus, in CR experiments, the usual assumption of no interference (Cox, 1958; Rubin, 1990) applies *only* at the cluster level. Moreover, in MPCR, assuming no interference only between pairs of clusters is sufficient. This advantage of MPCR designs can be substantial if contagion or social influence is present at the individual level, where, for example, individuals may affect the behavior of neighbors or friends, but such interference does not exist across clusters or pairs of clusters. Thus, we only assume:

ASSUMPTION 1 (No interference between matched-pairs). *Let $Y_{ijk}(\mathbf{T})$ be the potential outcomes for the $i$th unit in the $j$th cluster of the $k$th matched-pair where $\mathbf{T}$ is a $(m \times 2)$ matrix whose $(j, k)$ element is $T_{jk}$. We assume that if $T_{jk} = T'_{jk}$, then $Y_{ijk}(\mathbf{T}) = Y_{ijk}(\mathbf{T}')$.*

The assumption allows us to write $Y_{ijk}(T_{jk})$ rather than $Y_{ijk}(\mathbf{T})$. Since $T_{1k} = Z_k$ and $T_{2k} = 1 - Z_k$, $Y_{ijk}(T_{jk})$ only depends on $Z_k$. Given that the assumption of no interference among individuals is often highly unrealistic (Sobel, 2006), MPCR offers an attractive alternative. In the Mexico experiment, Assumption 1 is reasonable because most of the clusters in our experiment are noncontiguous and the travel times between them are substantial. However, especially in small villages, individual-level no interference assumptions would have been implausible.

Finally, we formalize the cluster-level randomized treatment assignment as follows.

ASSUMPTION 2 (Cluster randomization under matched-pair design). *The potential outcomes are independent of the randomization indicator variable: $(Y_{ijk}(1), Y_{ijk}(0)) \perp\!\!\!\perp Z_k$, for all $i, j$ and $k$. Also, $Z_k$ is independent across matched-pairs, and $\Pr(Z_k) = 1/2$ for all $k$.*

The assumption also implies $(Y_{ijk}(1), Y_{ijk}(0)) \perp\!\!\!\perp T_{jk}$ since $T_{jk}$ is a function of $Z_k$.

### 3.3 Quantities of Interest

We now offer the definitions of the causal effects of interest under MPCR (or CR in general) which have not been formally defined in the literature. At least two types of each of four distinct quantities may be of interest in these experiments. We begin with the four quantities, which define the target population, and then discuss the two types, which clarify the role of interference. Section 6 introduces additional quantities of interest when individual-level noncompliance exists. (All the quantities below are based on causal effects defined as grouped individual-level phenomena; we discuss cluster-level causal quantities in Section 4.6.)

3.3.1 *Target population quantities*. Table 1 offers an overview of the four target population causal effects. All four quantities represent the causal treatment effect (the potential outcome under treatment minus the potential outcome under control) averaged over different sets of units.

First is the *sample average treatment effect* (SATE or $\psi_S$), which is an average over the set of all units in the observed sample (which we denote as $\mathcal{S}$):

$$
\begin{aligned}
\psi_S &\equiv \mathbb{E}_{\mathcal{S}}(Y(1) - Y(0)) \\
&= \frac{1}{n} \sum_{k=1}^{m} \sum_{j=1}^{2} \sum_{i=1}^{n_{jk}} (Y_{ijk}(1) - Y_{ijk}(0)),
\end{aligned}
\tag{1}
$$

where the sums go over pairs, the two clusters within each pair and the units within each cluster.

The second quantity treats observed clusters as fixed (and not necessarily representative of some population) and the units within clusters as randomly sampled from the finite population of units within each cluster. This gives the *cluster average treatment effect* (CATE or $\psi_C$):

$$
\begin{aligned}
\psi_C &\equiv \mathbb{E}_{\mathcal{C}}(Y(1) - (0)) \\
&= \frac{1}{N} \sum_{k=1}^{m} \sum_{j=1}^{2} \sum_{i=1}^{N_{jk}} (Y_{ijk}(1) - Y_{ijk}(0)),
\end{aligned}
\tag{2}
$$

TABLE 1
*Quantities of Interest: For each causal effect, this table lists whether clusters and units within clusters are treated as observed and fixed or instead as a sample from a larger population. The resulting inferential target is also given*

| Quantities | | Clusters | Units within clusters | Inferential target |
|---|---|---|---|---|
| $\psi_S$ | SATE | Observed | Observed | Observed sample |
| $\psi_C$ | CATE | Observed | Sampled | Population within observed clusters |
| $\psi_U$ | UATE | Sampled | Observed | Observable units within the population of clusters |
| $\psi_P$ | PATE | Sampled | Sampled | Population |

where the expectation is taken over the set $\mathcal{C}$ which contains all observed units within the sample clusters, $N_{jk}$ is the known (and finite) population cluster size, and $N \equiv \sum_{k=1}^{m} (N_{1k} + N_{2k})$. Throughout, we assume simple random sampling within each cluster for simplicity, but other random sampling procedures can easily be accommodated via unit-level weights. Thus, the only difference between SATE and CATE is whether each unit within clusters is treated as fixed or randomly drawn based on a known sampling mechanism.

A third quantity treats the clusters as randomly sampled from a larger population, but the units within the sampled clusters are treated as fixed. The inferential target is the set $\mathcal{U}$, which includes all units in the population of clusters that would be observed if its cluster were in the observed sample. This is what we call the *unit average treatment effect* (UATE or $\psi_U$) and is defined as $\psi_U \equiv \mathbb{E}_{\mathcal{U}}(Y(1) - Y(0))$.

The final quantity of interest is the *population average treatment effect* (PATE or $\psi_P$), which is defined as $\psi_P \equiv \mathbb{E}_{\mathcal{P}}(Y(1) - Y(0))$, where the expectation is taken over the entire population $\mathcal{P}$—that is, the population of units within the population of clusters. For simplicity throughout, we assume an infinite population of clusters, but this is easily extended to finite populations at some cost in additional notation.

Researchers should design their experiments to make inferences to their desired quantity of interest, though in practice they may choose to estimate other quantities of interest when they face design limitations. In the SPS evaluation, for example, we would like to infer PATE for all of Mexico, but our health clusters were not (and due to political and administrative constraints could not be) randomly selected. This means that, like most medical experiments, any method applied to our data to estimate PATE will be dependent on assumptions about the selection process. An alternative approach would be to try to estimate one of the other quantities. CATE or SATE are straightforward possibilities, and CATE is probably most apt in this case, since individuals within clusters were randomly selected, and both quantities condition on the clusters we observe. Of course, even when inferences are made to restricted populations, readers may still extrapolate to a different population of interest, and so the researcher needs to decide on the appropriate presentation strategy. From a public policy perspective, UATE may be a reasonable target quantity, where we try to infer to the individuals who would be sampled in all the health clusters in Mexico that are similar to our observed clusters, and from which our clusters could plausibly have been randomly drawn.

3.3.2 *Interference.* Inference in CR experiments may be affected by three different types of interference, each of which may require different assumptions. First, when interference exists among individuals within a cluster, the potential outcomes of one person (or unit) within a cluster may be different depending on other units' treatment assignment. This type of interference is expected and no assumptions are required for the four causal quantities of interest. In CR experiments, within-cluster interference is part of the outcome, and researchers can estimate the causal effects of cluster-level treatment on unit-level outcome. Understanding the effect of individuals independent of and isolated from other individuals in the same cluster is best left to studies where individual randomization is possible.

Second, interference between clusters in *different* pairs may affect outcomes. Assumption 1 requires the absence of such interference between clusters in different pairs. We continue to maintain this assumption, as Sobel (2006) demonstrates that without it even the definition of a causal effect is complicated (see also Rosenbaum, 2007).

Third, interference between treatment and control clusters in the same pair requires us to redefine causal effects to account for interference. For example, if one cluster is assigned SPS, individuals in the other (control) cluster within the pair may become envious or depressed as a consequence. This type of interference within a pair can be dealt with in two ways. In the first, which we call *no-interference*, we define the causal effect (SATE, CATE, UATE or PATE) so that the treatment in one cluster has no effect on the potential outcomes of units in the control cluster. In the second, which we call the *with-interference*, the causal effect is defined so that it includes interference between clusters within pairs as well as interference between units within each cluster. (For our Mexico experiment, we do not expect much direct interference within or across pairs, although nearby clusters outside our experiment might exert some influence over those we observe, in which case the definition of UATE or PATE might change).

Estimating the no-interference version of SATE, CATE, UATE or PATE in the presence of interference is feasible only with assumption-laden estimators. In contrast, estimating the with-interference version is easier since it accepts whatever level of non-interference one's data happens to present. Of course, having a quantity that is easy to estimate is not a satisfactory substitute for having an estimate of the quantity of interest. The best way to avoid this problem is to use these facts to design better experiments. For example, we can select noncontiguous clusters to pair, and pairs that are not contiguous to other pairs. Following rules like this whenever feasible reduces the difference between the no-interference and with-interference quantities.

## 4. ESTIMATORS

We now define our estimators and derive their statistical properties. Our strategy throughout is to make as few assumptions as feasible beyond the experimental design. We also briefly discuss an approach that has been offered in the literature. Since our approach has little or no bias, and the existing estimator is biased but may have low variance in some circumstances, we also offer a model-based method that combines some of the benefits of both approaches.

### 4.1 Definitions

The point estimators for the with-interference version of the four quantities of interest are each weighted averages of within-pair mean differences between the treated and control clusters, but with different weights. We thus define

$$
\begin{aligned}
(3) \quad & \hat{\psi}(w_k) \\
& \equiv \frac{1}{\sum_{k=1}^{m} w_k} \\
& \cdot \sum_{k=1}^{m} w_k \Bigg\{ Z_k \Bigg( \frac{\sum_{i=1}^{n_{1k}} Y_{i1k}}{n_{1k}} - \frac{\sum_{i=1}^{n_{2k}} Y_{i2k}}{n_{2k}} \Bigg) \\
& \qquad + (1 - Z_k) \\
& \qquad\quad \cdot \Bigg( \frac{\sum_{i=1}^{n_{2k}} Y_{i2k}}{n_{2k}} - \frac{\sum_{i=1}^{n_{1k}} Y_{i1k}}{n_{1k}} \Bigg) \Bigg\},
\end{aligned}
$$

where the weight for the $k$th pair of clusters, denoted by $w_k$, defines a specific estimator.

The estimator most commonly recommended in the methodological literature is based on a weight using the harmonic mean of sample cluster sizes, which can be written as $\hat{\psi}(n_{1k} n_{2k}/(n_{1k} + n_{2k}))$ (see, e.g., Donner, 1987; Donner and Donald, 1987; Donner and Klar, 1993; Hayes and Bennett, 1999; Bloom, 2006; Raudenbush, 1997; Turner, White and Croudace, 2007). This estimator, and its variance estimator, are in general biased (see Appendix A.4), but may have low variance in some situations, an issue we return to in Section 4.5.

As shown in Table 2, $\hat{\psi}(n_{1k} + n_{2k})$ is our point estimator for both SATE and UATE, whereas $\hat{\psi}(N_{1k} + N_{2k})$ applies to both CATE and PATE. This is intuitive, as SATE and UATE are based on those units (which would be) sampled in a cluster whereas CATE and PATE are based on the population of units within clusters. Our estimator for SATE and UATE differs from the existing estimator based on harmonic mean weights unless the sample cluster sizes within each matched pair are equal ($n_{1k} = n_{2k}$ for all $k = 1, \ldots, m$), which rarely occurs at least in field experiments.

Table 2 also summarizes the variances and their estimators. Under our design-based inference, UATE and PATE have identifiable variances, the exact expression for which we give below. SATE and CATE have unidentifiable variances, and so we offer their upper bound, leading to a conservative confidence interval. Our variance estimators differ from the existing estimator even when sample cluster sizes are matched exactly. Our variance estimator is approximately unbiased for any weights. Estimates from UATE and PATE (or equivalently SATE and CATE) will differ depending on how sample and population sizes vary across clusters.

## 4.2 Bias

We first focus on SATE. This allows us, following Neyman (1923), to use the randomized treatment assignment mechanism as the sole basis for statistical inference (see also Imai, 2008). Here, the potential outcomes are assumed fixed, but possibly unknown, quantities. We begin by rewriting $\hat{\psi}(n_{1k} + n_{2k})$ using potential outcome notation:

$$
\begin{aligned}
& \hat{\psi}(n_{1k} + n_{2k}) \\
& = \frac{1}{n} \sum_{k=1}^{m} (n_{1k} + n_{2k}) \\
& \quad \cdot \Bigg\{ Z_k \Bigg( \frac{\sum_{i=1}^{n_{1k}} Y_{i1k}(1)}{n_{1k}} - \frac{\sum_{i=1}^{n_{2k}} Y_{i2k}(0)}{n_{2k}} \Bigg) \\
& \qquad + (1 - Z_k) \\
& \qquad\quad \cdot \Bigg( \frac{\sum_{i=1}^{n_{2k}} Y_{i2k}(1)}{n_{2k}} - \frac{\sum_{i=1}^{n_{1k}} Y_{i1k}(0)}{n_{1k}} \Bigg) \Bigg\}.
\end{aligned}
$$

Then, taking the expectation with respect to $Z_k$ yields

$$
\begin{aligned}
(4) \quad & E_a\{\hat{\psi}(n_{1k} + n_{2k})\} - \psi_S \\
& = \frac{1}{n} \sum_{k=1}^{m} \sum_{j=1}^{2} \Bigg\{ \Bigg( \frac{n_{1k} + n_{2k}}{2} - n_{jk} \Bigg) \\
& \qquad\quad \cdot \sum_{i=1}^{n_{jk}} \frac{Y_{ijk}(1) - Y_{ijk}(0)}{n_{jk}} \Bigg\},
\end{aligned}
$$

where the expectation is taken with respect to the randomization of treatment assignment which we indicate by the subscript "$a$."

Although the bias does not generally equal zero, either of two common conditions can eliminate it. These two conditions motivate our choice of weights ($w_k = n_{1k} + n_{2k}$). First, when cluster sizes are equal within each matched-pair (i.e., $n_{1k} = n_{2k}$ for all $k$), the bias is always zero. This implies that researchers may wish to form pairs of clusters, at least partially, based on their sample cluster size if SATE is the estimand. Second, $\hat{\psi}(n_{1k} + n_{2k})$ is also unbiased if matching is effective, so that the within-cluster SATEs are identical for each matched-pair (i.e., $\sum_{i=1}^{n_{1k}}(Y_{i1k}(1) - Y_{i1k}(0))/n_{1k} = \sum_{i=1}^{n_{2k}}(Y_{i2k}(1) - Y_{i2k}(0))/n_{2k}$ for all $k$). In contrast, bias may remain if cluster sizes are poorly matched *and* within each pair cluster sizes are strongly associated with the cluster-specific SATEs. However, the bounds on the bias can be found by applying the Cauchy–Schwarz inequality to equation (4) and they can be consistently estimated from the observed data. In sum, roughly speaking, if cluster sizes and important confounders are matched well so that pre-randomization matching accomplishes the purpose for which it was designed, this estimator will be approximately unbiased.

TABLE 2
*Point estimators and variances for the four causal quantities of interest. "Identified" refers to design-based identification of estimated causal effects without modeling assumptions*

|                | **SATE** | **CATE** | **UATE** | **PATE** |
|----------------|----------|----------|----------|----------|
| Point estimator | $\hat{\psi}(n_{1k} + n_{2k})$ | $\hat{\psi}(N_{1k} + N_{2k})$ | $\hat{\psi}(n_{1k} + n_{2k})$ | $\hat{\psi}(N_{1k} + N_{2k})$ |
| Variance       | $\mathrm{Var}_a(\hat{\psi})$ | $\mathrm{Var}_{au}(\hat{\psi})$ | $\mathrm{Var}_{ap}(\hat{\psi})$ | $\mathrm{Var}_{aup}(\hat{\psi})$ |
| Identified     | no       | no       | YES      | YES      |

A similar bias expression can be derived for our CATE estimator, $\hat{\psi}(N_{1k} + N_{2k})$, where the weights are now based on the arithmetic mean of the population cluster sizes rather than their sample counterparts. A calculation analogous to the one above yields the following bias expression:

$$
\begin{aligned}
(5) \quad & E_{au}(\hat{\psi}(N_{1k} + N_{2k})) - \psi_C \\
& = \frac{1}{N} \sum_{k=1}^{m} \sum_{j=1}^{2} \left\{ \left( \frac{N_{1k} + N_{2k}}{2} - N_{jk} \right) \right. \\
& \qquad \left. \cdot E_u(Y_{ijk}(1) - Y_{ijk}(0)) \right\},
\end{aligned}
$$

where subscript "$au$" means that the expectation is taken with respect to random treatment assignment and the simple random sampling of units within each cluster. The conditions under which this bias disappears are analogous to the ones for SATE: If matching is effective so that the cluster-specific average causal effects, that is, $E_u[Y_{ijk}(1) - Y_{ijk}(0)]$, are constant across clusters within each pair, then the bias is zero. The bias also vanishes if the population cluster sizes are identical within each pair, that is, $N_{1k} = N_{2k}$ for all $k$. Again, the bounds on the bias can be obtained in the manner similar to the case of SATE above.

Finally, the bias for UATE and PATE can be obtained by taking the expectation of the bias for SATE and CATE, respectively, with the expectation defined with based on random sampling of cluster pairs. If the within-cluster sample (population) average treatment effects are uncorrelated with cluster sizes within each matched-pair, then the bias for the estimation of UATE (PATE) is zero, regardless of whether one can match exactly on cluster sizes. In general, however, cluster sizes may be correlated with the size of average treatment effects. In such cases, the matching strategies to reduce the bias for the estimation of SATE (CATE) also work for the estimation of UATE (PATE). That is, pairs of clusters should be constructed such that within each pair, cluster sizes and important pre-treatment covariates are similar. (We also derived an unbiased estimator and its variance, but we do not present it here because they are not invariant to a constant shift of the outcome variable when cluster sizes vary within each pair.)

### 4.3 Variance

In a critical comment about Klar and Donner (1997), Thompson (1998) shows how to obtain valid variance estimates under the linear mixed effects model and the "common effect assumption." In their reply, Klar and Donner (1998) criticize the common effect assumption and, as a result, maintain their claim of analytical difficulties with MPCRs. We show here how to obtain valid variance estimates without the common treatment effect assumption or other modeling assumptions.

Rather than focusing on each of our proposed estimators, $\hat{\psi}(n_{1k} + n_{2k})$ and $\hat{\psi}(N_{1k} + N_{2k})$, separately we consider the variance of the general estimator, $\hat{\psi}(w_k)$ in equation (3), so that the analytical results we develop apply to any choice of weights including the harmonic mean weights. For notational simplicity, we use normalized weights, that is, $\tilde{w}_k \equiv nw_k / \sum_{k=1}^{m} w_k$ (so that the weights sum up to $n$ as in our estimator of SATE and UATE), and consider the variances of $\hat{\psi}(\tilde{w}_k)$. First, we use potential outcomes notation and write $\hat{\psi}(\tilde{w}_k) = \sum_{k=1}^{m} \tilde{w}_k \{Z_k D_k(1) + (1 - Z_k) D_k(0)\} / n$. Then, our variance estimator is

$$
\begin{aligned}
& \hat{\sigma}(\tilde{w}_k) \\
& \equiv \frac{m}{(m-1)n^2}
\end{aligned}
$$

$$
\begin{aligned}
&\cdot \sum_{k=1}^{m} \Bigg[ \tilde{w}_k \bigg\{ Z_k \bigg( \frac{\sum_{i=1}^{n_{1k}} Y_{i1k}}{n_{1k}} - \frac{\sum_{i=1}^{n_{2k}} Y_{i2k}}{n_{2k}} \bigg) \\
&\quad + (1 - Z_k) \\
&\qquad \cdot \bigg( \frac{\sum_{i=1}^{n_{2k}} Y_{i2k}}{n_{2k}} - \frac{\sum_{i=1}^{n_{1k}} Y_{i1k}}{n_{1k}} \bigg) \bigg\} \\
&\qquad\qquad - \frac{n \hat{\psi}(\tilde{w}_k)}{m} \Bigg]^2 .
\end{aligned}
\tag{6}
$$

*SATE.* We first consider the variance of $\hat{\psi}(\tilde{w}_k)$ for SATE. Taking the expectation of $\hat{\psi}(\tilde{w}_k)$ with respect to $Z_k$, the true variance of $\hat{\psi}(\tilde{w}_k)$ is given by

$$
\operatorname{Var}_a(\hat{\psi}(\tilde{w}_k)) = \frac{1}{4n^2} \sum_{k=1}^{m} \tilde{w}_k^2 (D_k(1) - D_k(0))^2 .
\tag{7}
$$

This variance is not identified since we do not jointly observe $D_k(1)$ and $D_k(0)$ for each $k$. Thus, we identify an upper bound of this variance, making no additional assumptions, and estimate it from the observed data.

The next proposition establishes that the true variance, $\operatorname{Var}_a(\hat{\psi}(\tilde{w}_k))$, is not identifiable, and shows that our proposed variance estimator, $\hat{\sigma}(\tilde{w}_k)$, is conservative.

PROPOSITION 1 (SATE variance identification). *Suppose that SATE is the estimand. Then, the true variance of $\hat{\psi}(\tilde{w}_k)$ is not identifiable. The bias of $\hat{\sigma}(\tilde{w}_k)$ is given by*

$$
\begin{aligned}
&E_a(\hat{\sigma}(\tilde{w}_k)) - \operatorname{Var}_a(\hat{\psi}(\tilde{w}_k)) \\
&\quad = \frac{m}{4n^2} \operatorname{var}\{\tilde{w}_k (D_k(1) + D_k(0))\},
\end{aligned}
$$

*where $\operatorname{var}(\cdot)$ represents the sample variance with denominator $m - 1$.*

See Appendix A.1 for a proof. This proposition implies that on average $\hat{\sigma}(\tilde{w}_k)$ overestimates the true variance $\operatorname{Var}_a(\hat{\psi}(\tilde{w}_k))$ unless the sample variance of weighted within-cluster SATEs across pairs is zero. For example, if SATE is constant across pairs, and the cluster sizes are equal, $\hat{\sigma}(\tilde{w}_k)$ estimates the true variance without bias. However, such a scenario is highly unlikely under MPCR, and thus $\hat{\sigma}(\tilde{w}_k)$ should be seen as a conservative estimator of the variance. It is also possible to obtain a less conservative variance estimate than $\hat{\sigma}(\tilde{w}_k)$. For example, researchers may use a consistent estimator of $\{(\sum_{k=1}^{m} \tilde{w}_k^2 D_k(1)^2)^{1/2} + (\sum_{k=1}^{m} \tilde{w}_k^2 D_k(1)^2)^{-1/2}\}^2 / 4n^2$, which is obtained by applying the Cauchy–Schwarz inequality to equation (7). Another approach to obtain a tighter bound would be to apply the covariance inequality to the bias expression given in Proposition 1.

*CATE.* Next, we study variance for CATE, $\hat{\psi}(\tilde{w}_k)$, which we write as

$$
\begin{aligned}
&\operatorname{Var}_{au}(\hat{\psi}(\tilde{w}_k)) \\
&\quad = E_u\{\operatorname{Var}_a(\hat{\psi}(\tilde{w}_k))\} \\
&\qquad + \operatorname{Var}_u\{E_a(\hat{\psi}(\tilde{w}_k))\} \\
&\quad = \frac{1}{4n^2} \sum_{k=1}^{m} \tilde{w}_k^2 \{E_u(D_k(1) - D_k(0))^2 \\
&\qquad + \operatorname{Var}_u(D_k(1) + D_k(0))\},
\end{aligned}
\tag{8}
$$

where the second equality holds because sampling of units is independent within clusters. Similar to the SATE variance, this is not identified since we do not jointly observe $D_k(1)$ and $D_k(0)$ for each $k$. The next proposition shows that $\hat{\sigma}(\tilde{w}_k)$ is again conservative.

PROPOSITION 2 (CATE variance identification). *Suppose that CATE is the estimand. The true variance of $\hat{\psi}(\tilde{w}_k)$, $\operatorname{Var}_{au}(\hat{\psi}(\tilde{w}_k))$, is not identifiable. The bias of $\hat{\sigma}(\tilde{w}_k)$ is given by*

$$
\begin{aligned}
&E_a(\hat{\sigma}(\tilde{w}_k)) - \operatorname{Var}_a(\hat{\psi}(\tilde{w}_k)) \\
&\quad = \frac{m}{4n^2} \operatorname{var}\{\tilde{w}_k E_u(D_k(1) + D_k(0))\}.
\end{aligned}
$$

See Appendix A.2 for a proof. The proposition implies that our proposed variance estimator, $\hat{\sigma}(\tilde{w}_k)$, is an upper bound of the true variance. As in the case of the SATE, this upper bound can be improved. For example, rewrite the variance in equation (8) as

$$
\begin{aligned}
&\operatorname{Var}_{au}(\hat{\psi}(\tilde{w}_k)) \\
&\quad = \frac{1}{2n^2} \sum_{k=1}^{m} \tilde{w}_k^2 \Bigg[ \operatorname{Var}_u(D_k(1)) + \operatorname{Var}_u(D_k(0)) \\
&\qquad + \frac{1}{2}\{E_u(D_k(1) - D_k(0))\}^2 \Bigg].
\end{aligned}
\tag{9}
$$

Then, apply the Cauchy–Schwarz inequality to the third term in the bracket of equation (9). Alternatively, applying the covariance inequality to the bias expression in Proposition 2 yields a tighter bound.

*UATE and PATE.* Unlike in the case of the SATE and the CATE, the variance of $\hat{\psi}$ is identified and can estimated approximately without bias when UATE or PATE is the estimand. We establish this result as the following proposition:

PROPOSITION 3. *Conditional on $\bar{w} = \sum_{k=1}^{m} w_k / m$, the variances of $\hat{\psi}(\tilde{w}_k)$ for estimating the UATE and PATE are given by:*

$$\mathrm{Var}_{ap}(\hat{\psi}(\tilde{w}_k)) = \frac{1}{m\bar{w}^2} \mathrm{Var}_p(w_k D_k),$$

$$\mathrm{Var}_{apu}(\hat{\psi}(\tilde{w}_k)) = \frac{1}{m\bar{w}^2}[E_p\{w_k^2 \mathrm{Var}_u(D_k)\} + \mathrm{Var}_p\{\tilde{w}_k E_u(D_k)\}],$$

*respectively, where $D_k = Z_k D_k(1) + (1 - Z_k)D_k(0)$ and "$p$" represents the expectation with respect to simple random sampling of matched-pairs of clusters. Conditional on $\bar{w}$, both variances can be estimated by $\hat{\sigma}(\tilde{w}_k)$ without bias under their corresponding sampling schemes.*

See Appendix A.3 for a proof. The proposition shows that when estimating PATE, the variance of $\hat{\psi}(\tilde{w}_k)$ is proportional to the sum of two elements: the mean of within-cluster variances and the variance of within-cluster means. If all units are included in each cluster, then the first term will be zero because the within-cluster means are observed without sampling uncertainty, that is, $\mathrm{Var}_u(D_k) = 0$ for all $k$. In either case, however, our proposed variance estimator $\hat{\sigma}(\tilde{w}_k)$ is unbiased, conditional on the mean weight, $\bar{w}$.

*Inference.* Given our proposed estimators and variances, we make statistical inferences by assuming that $\hat{\psi}(\tilde{w}_k)$ is approximately unbiased. We consider three situations:

1. *Many pairs.* When the number of pairs is large (regardless of the number of units within each cluster), no additional assumption is necessary due to the central limit theorem. For PATE and UATE, the level $\alpha$ confidence intervals are given by $[\hat{\psi}(\tilde{w}_k) - z_{\alpha/2}\sqrt{\hat{\sigma}(\tilde{w}_k)}, \hat{\psi}(\tilde{w}_k) + z_{\alpha/2}\sqrt{\hat{\sigma}(\tilde{w}_k)}]$ where $z_{\alpha/2}$ represents the critical value of two-sided level $\alpha$ normal test. For the SATE and CATE, the confidence level of this interval will be greater than or equal to $\alpha$.

2. *Few pairs, many units.* For CATE (and PATE), the central limit theorem implies that $D_k$ follows the normal distribution. Since the weights are assumed to be fixed for CATE, $\tilde{w}_k D_k$ is also normally distributed. For the other three quantities, we assume $\tilde{w}_k D_k$ is normally distributed. In either case, the level $\alpha$ confidence intervals are given by $[\hat{\psi}(\tilde{w}_k) - t_{m-1,\alpha/2}\sqrt{\hat{\sigma}(\tilde{w}_k)}, \hat{\psi}(\tilde{w}_k) + t_{m-1,\alpha/2}\sqrt{\hat{\sigma}(\tilde{w}_k)}]$, where $t_{m-1,\alpha/2}$ represents the critical value of the one-sample, two-sided level $\alpha$ $t$-test with $(m - 1)$ degrees of freedom. For the SATE and CATE, the confidence level of this interval will be greater than or equal to $\alpha$.

3. *Few pairs, few units.* When little information is available, a distributional assumption is required for the inferences about all four quantities. We may assume $\tilde{w}_k D_k$ follows the normal distribution as above and construct the confidence intervals and conduct hypothesis tests based on $t$-distribution.

Finally, although it was once thought that the need for, and inability to estimate, the intracluster correlation coefficient (ICC) was a major disadvantage of MPCR designs (Campbell, Mollison and Grimshaw, 2001; Klar and Donner, 1997; Donner, 1998), estimates of the ICC are in fact not needed for our estimators or their variances. Below, we also show that efficiency analysis, power comparisons and sample size calculations can also be conducted without the ICC estimation.

### 4.4 Performance in Practice

We now study how our estimator and the harmonic mean estimator work in practice. The results here also motivate a combined approach to estimation we offer in Section 4.5.

*Confidence interval coverage.* To construct realistic simulations, we begin with the observed cluster-specific mean for two out-of-pocket health expenditures from the SPS evaluation data (measured in pesos) and use this to set the potential outcomes' true population for the simulation. Finally, we generate the outcome variables via independent normal draws for units within clusters using a set of heterogeneous variances. Thus, the existing harmonic mean estimator's mean and variance constancy assumptions are violated, as is common in real data, although its normality and independence assumptions are maintained. (Replication data are available in Imai, King and Nall, 2009.)

We study the properties of the proposed and existing variance estimators with PATE or UATE as the estimand. (As shown in Proposition 2, the CATE variance is not identified and the expectation of our variance estimator equals a upper bound.) We generate a population of clusters by bootstrapping the observed pairs of SPS clusters along with their observed means and a set of heterogeneous variances. We then compute coverage probabilities under both estimators where the arithmetic and harmonic mean weights are used for the proposed and existing estimators, respectively. We draw from the discrete empirical distribution, which is far from a Gaussian distribution, yielding a hard case for both estimators. The left panels of Figure 1 summarize the results. As expected due to the central limit theorem, both sets of our 90% confidence intervals (solid disks) approach their corresponding nominal coverage probabilities as the number of pairs increase. In contrast, the confidence intervals based on the harmonic mean variance estimator (open diamonds) are biased—too wide in the top graph and too narrow in the bottom—and the magnitude of bias does not decrease even as the number of pairs grows.

*Standard error comparisons.* We begin by computing the standard error (the square root of the estimated variance) based on the general variance formula proposed in Donner (1987), Donner and Donald (1987) and Donner and Klar (1993), as well as the one based on our approximately unbiased alternative. For comparability, we use the arithmetic mean weights for both standard error calculations. We make these computations for a large number of outcome variables from the SPS evaluation survey conducted 10 months after randomization. The outcome variables include some which are binary (e.g., did the respondent suffer catastrophic medical expenditures? Does our blood test indicate that the respondent has high cholesterol? Has the respondent been diagnosed with asthma?) and others denominated in Mexican pesos (e.g., out-of-pocket expenditures for health care, for drugs, etc.). We then divided this standard error by our alternative for each variable. The top right graph in Figure 1 gives a smoothed histogram of these ratios (plotted on the log scale but labeled in original units, with 1 the point of equality). In these real data, the biased standard errors range from about two times too small to two times too large. Note that the central tendency of this histogram has no particular meaning, as it is constructed from whatever questions happened to be asked on the survey. The key point is that in real data the deviation from the approximately unbiased estimator for any *one* such standard error can be large in either direction.

*Bias-variance tradeoff.* Using data from an expenditure outcome in the SPS sample, we simulate an instance in which the variance of the existing estimator outperforms our estimator. To distinguish between the harmonic and arithmetic mean, we begin by setting all within-pair cluster sizes equal to the size of the treatment cluster in the SPS evaluation. Then, keeping the total pair size constant, we increase the difference in within pair cluster size such that the added difference in cluster sizes is proportional to the within-pair treatment effect. This leaves the average treatment effect constant while demonstrating differences in the two weighting schemes. The bottom right graph in Figure 1 presents the absolute difference between the two estimators in mean square error, squared bias and variance, with the observed SPS value marked with a vertical line.

The overall picture from these results indicates that the arithmetic estimator would be preferred because it has lower mean square error than the harmonic mean estimator. However, at the expense of introducing bias when treatment effect is both variable across pairs and correlated with the cluster size, the harmonic mean estimator can have substantially lower variance. These results suggest the possibility of an improved estimator based on the combination of both approaches, a subject to which we now turn.

### 4.5 An Encompassing, Model-Based Approach

The standard harmonic mean estimator is unbiased when applied to data where the between-cluster homogeneity assumption holds. In this situation, the harmonic mean weights also have the attractive property of downweighting observations with worse matches and larger variances, thereby reducing variance. If the homogeneity assumption is violated, however, then one cannot afford to downweight pairs, no matter how badly matched or imprecisely measured, because doing so could result in arbitrarily large biases. In contrast, the arithmetic mean estimator avoids bias by making no assumptions about the nature of how treatment effects vary over the pairs. However, a consequence of it imposing no structure on treatment effect heterogeneity is that mismatched pairs are not downweighted and so some inefficiency may result if in fact the treatment effects are similar across pairs.

We now combine the insights of these two approaches and propose a single encompassing model that provides some of the advantages of each, at the cost of somewhat more stringent assumptions than with our design-based approach. Consider data with $m^*$ groups of clusters, where the homogeneity assumption holds within each group. Assume that the clusters within any one pair are never split between groups. Let $g(k) = l$ denote the group to which pair $k$ belongs, $l = 1, 2, \ldots, m^*$ with $m^* \leq m$. Then, make the modeling assumption that $Y_{ijk}(t) \overset{\text{i.i.d.}}{\sim} \mathcal{N}(\mu_t^{(g(k))}, \tilde{\sigma}^{(g(k))})$ for $t = 0, 1$ where $\mu_t^{(l)}$ and $\tilde{\sigma}^{(l)}$ are not necessarily equal to $\mu_t^{(l')}$ and $\tilde{\sigma}^{(l')}$ for $l \neq l'$. Under this model, CATE equals,

$$(10) \quad \psi_C = \frac{1}{N} \sum_{k=1}^{m} (N_{1k} + N_{2k})(\mu_1^{(g(k))} - \mu_0^{(g(k))}).$$

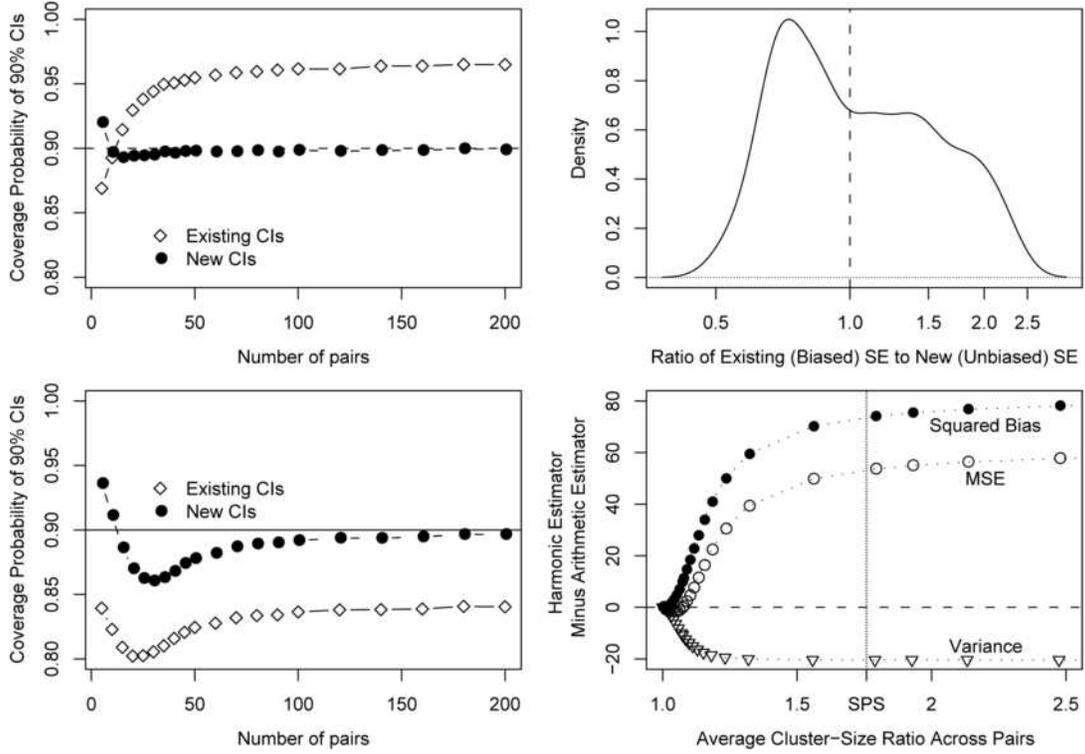

Fig. 1. *Inference Accuracy. Simulations in the left panels demonstrate how our estimator's coverage is approximately correct and increasingly so for larger sample sizes, while the existing estimator can yield confidence intervals that are either too large (top left) or too small (bottom left). The top right panel uses real data to give the ratios of the harmonic mean standard error to our approximately unbiased alternative on the horizontal axis (on the log scale, but labeled as ratios). The bottom right figure gives squared bias, MSE, and variance comparisons as a function of the average cluster size ratio; a vertical line marks the observed heterogeneity in the SPS data.*

When the group membership is known ex ante, an unbiased and efficient estimator of CATE is given by replacing $\mu_k^{(l)} \equiv \mu_1^{(l)} - \mu_0^{(l)}$ with its harmonic mean estimate

$$\hat{\mu}_k^{(l)} = \sum_{k=1}^{m} \mathbf{1}\{g(k) = l\} w_k D_k \bigg/ \left[ \sum_{k'=1}^{m} \mathbf{1}\{g(k') = l\} w_{k'} \right],$$

where $w_k = n_{1k} n_{2k}/(n_{1k} + n_{2k})$ and $\mathbf{1}\{\cdot\}$ represents the indicator function. Thus, this mixture model estimator is an arithmetic mean of within (homogeneous) group harmonic mean estimators. A special case is the harmonic mean estimator in the literature, where the homogeneity assumption is made across all clusters, that is, $m^* = 1$. When every pair belongs to a different group, that is, $m^* = m$, this estimator approximates our proposed design-based estimator. In many applications, the group membership as well as the number of groups may be unknown. In this case, CATE may be estimated via standard methods for fitting finite mixture models (e.g., McLaughlan and Peel, 2000).

### 4.6 Cluster-Level Quantities of Interest

The eight quantities of interest defined in Section 3.3—SATE, CATE, PATE and UATE, both with and without interference—are all defined as aggregations of unit-level causal effects. For some purposes, however, analogous quantities of interest can be defined at the cluster level. For example, quantities of interest in the SPS evaluation include the health clinic-level variables. Some of these effects, such as the supply of drugs and doctors, are defined and measured at the health clinic, and so are effectively unit-level variables amenable to cluster-level analyses.

However, for other variables, individual-level survey responses are required to measure the aggregate variables. Examples include the success health clinics in our experiment have in protecting privacy, reduce waiting times, etc. If these latter variables are used to judge the causal effect of SPS on the clinics, we have a CR experiment, but a quantity of interest at the cluster level. In this situation, our estimator is a special case of equation (3), with a constant weight, $\hat{\psi}(1)$. Similarly, the variance of this estimator is a special case of our general formulation in equation (6), $\hat{\sigma}(1)$. This estimator for aggregate quantities is unbiased and invariant for all quantities of interest. In the case of unit-level variables amenable to cluster-level analysis (such as collected via survey), there will likely be sampling error and so may result in a larger variance.

## 5. COMPARING MATCHED-PAIR AND OTHER DESIGNS

We now study the relative efficiency and power of the MPCR and unmatched cluster randomization (UMCR) designs, and give sample size calculations for MPCR. We also briefly compare MPCR with the stratified design and discuss the consequences of loss of clusters under each.

### 5.1 Unmatched Cluster Randomized Design

The UMCR design is defined as follows. Consider a random sample of $2m$ clusters from a population. We observe a total of $n_j$ units within the $j$th cluster in the sample, and use $n$ to denote the total number of units in the sample, $n = \sum_{j=1}^{2m} n_j$. Under this design, $m$ randomly selected clusters are assigned to the treatment group with equal probability while the remaining $m$ clusters are assigned to the control group.

We construct an estimator analogous to that proposed for the UMCR as

$$
\begin{aligned}
\hat{\tau}(\tilde{w}_j) &\equiv \frac{2}{n} \sum_{j=1}^{2m} \sum_{i=1}^{n_j} \frac{\tilde{w}_j}{n_j} \{Z_j Y_{ij} - (1 - Z_j) Y_{ij}\} \\
&= \frac{2}{n} \sum_{j=1}^{2m} \sum_{i=1}^{n_j} \frac{\tilde{w}_j}{n_j} \{Z_j Y_{ij}(1) - (1 - Z_j) Y_{ij}(0)\},
\end{aligned}
$$

where $Z_j$ is the randomized binary treatment variable, $Y_{ij}(t)$ is the potential outcome for the $i$th unit in the $j$th cluster under the treatment value $t$ for $t = 0, 1$, and $\tilde{w}_j$ is the known normalized weight with $\sum_{j=1}^{2m} \tilde{w}_j = n$. For SATE and UATE, we use $\tilde{w}_j = n_j$. For CATE and PATE, we use $\tilde{w}_j \propto N_j$ where $N_j$ is the population size of the $j$th cluster. Analysis similar to the one in Section 4.2 shows that this estimator is unbiased for all four quantities in UMCR experiments.

The commonly used estimator in the literature for this design takes a form slightly different from equation (11): $\hat{\kappa} \equiv \sum_{j=1}^{2m} Z_j \sum_{i=1}^{n_j} Y_{ij} / \sum_{j=1}^{2m} Z_j n_j + \sum_{j=1}^{2m} (1 - Z_j) \sum_{i=1}^{n_j} Y_{ij} / (n - \sum_{j=1}^{2m} Z_j n_j)$. This estimator is applicable to SATE and UATE but not CATE and PATE because it ignores cluster population weights. The estimator is also biased for SATE and UATE, and the magnitude of bias can be derived using the Taylor series. Without modeling assumptions, the exact variance calculation is difficult within the design-based framework because the denominator as well as the numerator is a function of the randomized treatment variable. In addition, the usual approximate variance calculations for such a ratio estimator yield either the same variance as $\hat{\tau}(n_j)$ or the variance estimator that is not invariant to a constant shift. Thus, for the sake of simplicity, we focus on $\hat{\tau}(\tilde{w}_j)$ in this section although $\hat{\kappa}$ and its approximate variance estimator may perform reasonably well in practice.

For the rest of this section, we assume that the estimand is UATE. However, the same calculations apply when the estimand is PATE since the variance estimator is the same for both. For SATE and CATE, we can interpret these results as conservative estimates of efficiency, power and sample sizes.

### 5.2 Efficiency

When the estimand is UATE, the variance of $\hat{\tau}(\tilde{w}_j)$ is approximately (conditional on $n = \sum_{j=1}^{2m} \tilde{w}_j$) equal to $\mathrm{Var}_{ac}(\hat{\tau}(\tilde{w}_j)) = \frac{4m}{n^2} \{\mathrm{Var}_c(\tilde{w}_j \overline{Y_j(1)}) + \mathrm{Var}_c(\tilde{w}_j \cdot \overline{Y_j(0)})\}$, where $\overline{Y_j(t)} \equiv \sum_{i=1}^{n_j} Y_{ij}(t)/n_j$ for $t = 0, 1$, and the subscript "$c$" represents the simple random sampling of clusters. To facilitate comparison, assume that under MPCR one is able to match on cluster sizes so that $n_{1k} = n_{2k}$ for all $k$. Proposition 3 implies that under the same condition the variance of $\hat{\psi}(\tilde{w}_k)$ can be approximated by $\mathrm{Var}_{ap}(\hat{\psi}(\tilde{w}_k)) = m \mathrm{Var}_p\{\tilde{w}_k(\overline{Y_{jk}(1)} - \overline{Y_{j'k}(0)})\}/n^2$, where $\overline{Y_{jk}(t)} \equiv \sum_{i=1}^{n_{jk}} Y_{ijk}(t)/n_{jk}$ and $j \neq j'$. Since the assumption of $n_{1k} = n_{2k}$ means $\tilde{w}_{jk} = 2\tilde{w}_j$, we have $\mathrm{Var}_p(\tilde{w}_k \overline{Y_{jk}(t)}) =$ $4 \operatorname{Var}_c(\tilde{w}_j \overline{Y_j(t)})$ for $t = 0, 1$. Thus, the relative efficiency of MPCR over UMCR is

$$\frac{\operatorname{Var}_{ac}(\hat{\tau}(\tilde{w}_j))}{\operatorname{Var}_{ap}(\hat{\psi}(\tilde{w}_k))}$$
$$\approx \left\{ 1 - \frac{2 \operatorname{Cov}_p(\tilde{w}_k \overline{Y_{jk}(1)}, \tilde{w}_k \overline{Y_{j'k}(0)})}{\sum_{t=0}^1 \operatorname{Var}_p(\tilde{w}_k \overline{Y_{jk}(t)})} \right\}^{-1}.$$

This implies that the relative efficiency of MPCR depends on the correlation of the observed within-pair cluster mean outcomes weighted by cluster sizes. If matching induces a positive correlation, as is its purpose and will normally occur in practice, then MPCR is more efficient. (In the worst case scenario where matching is implemented in a manner opposite to the way it was designed, and thus induces a negative correlation, MPCR can be less efficient.) Under MPCR, we can estimate $\operatorname{Cov}_p(\tilde{w}_k \overline{Y_{jk}(1)}, \tilde{w}_k \overline{Y_{j'k}(0)})$ without bias using the sample covariance between $\tilde{w}_k \overline{Y_{jk}(1)}$ and $\tilde{w}_k \overline{Y_{j'k}(0)}$, which are jointly observed for each $k$. And thus, under MPCR, the variance one would obtain under UMCR can also be estimated without bias (see also Imai, 2008). This is another advantage of MPCR since the converse is not true. (If cluster sizes are equal, one can also estimate the ICC nonparametrically and separately for the treated and control groups—there is no reason to assume the ICC is the same for two potential outcomes as done in the literature. Note that the ICC is not required for efficiency, power or sample size calculations.)

*Empirical evidence.* Although the MPCR design have other advantages in public policy evaluations (King et al., 2007), their advantage in statistical efficiency can be considerable. We estimate the efficiency of MPCR as used in the SPS evaluation over the efficiency that our experiment would have achieved, if we had used complete randomization without matching. Figure 2 plots the relative efficiency of our estimator for MPCR over UMCR for UATE and for PATE. We do this for our 14 outcome variables denominated in pesos. For UATE, the estimator based on the MPCR is between 1.13 and 2.92 times more efficient, which means that our standard errors would have been as much as $\sqrt{2.92} = 1.7$ times larger if we had neglected to pair clusters first. The result is even more dramatic for estimating PATE, for which the MPCR design for different variables is between 1.8 and 38.3 times more efficient. In this situation, our standard errors would have been as much as six times larger if we had neglected to match first.

## 5.3 Power

We now use the variance results in Section 4.3 to calculate statistical power, that is, the probability of rejecting the null if it is indeed false, for UATE and PATE, which also represent the minimum power for SATE and CATE, respectively.

### 5.3.1 *Power calculations under the matched-pair design.*
We begin with power calculation for UATE given a null hypothesis of $H_0 : \psi_U = 0$, the alternative hypothesis of $H_A : \psi_U = \psi$, and the level $\alpha$ $t$-test. In this setting, Proposition 3 implies the power function, $1 + \mathcal{T}_{m-1}(-t_{m-1,\alpha/2} \mid n\psi/\sqrt{m \operatorname{Var}_p\{\tilde{w}_k D_k\}} - \mathcal{T}_{m-1}(t_{m-1,\alpha/2} \mid n\psi/\sqrt{m \operatorname{Var}_p\{\tilde{w}_k D_k\}})$, where $\mathcal{T}_{m-1}(\cdot \mid \zeta)$ is the distribution function of the noncentral $t$ distribution with $(m-1)$ degrees of freedom and the noncentrality parameter $\zeta$, and $\tilde{w}_k = n_{1k} + n_{2k}$. For UATE, we sample cluster pairs but not units within each cluster. Thus, a simpler expression for the power function results if we assume equal cluster sizes. In that case, a researcher may reparameterize the power function by normalizing $\psi$ in terms of the standard deviation of within-pair mean differences, that is, $d_U \equiv \psi/\sqrt{\operatorname{Var}(D_k)}$. Then, we write the power function as

$$1 + \mathcal{T}_{m-1}(-t_{m-1,\alpha/2} \mid d_U \sqrt{m}) \tag{12}$$
$$- \mathcal{T}_{m-1}(t_{m-1,\alpha/2} \mid d_U \sqrt{m}).$$

Next, for PATE, we sample units within each cluster as well as pairs of clusters. The null hypothesis is given by $H_0 : \psi_P = 0$ and the alternative is $H_a : \psi_P = \psi$. Again, for simplicity, we assume $\bar{n} = n_{jk} = n/(2m)$ for all $j$ and $k$. Then, Proposition 3 implies the power function is of the same form as equation (12) except that the noncentrality parameter is given by

$$\psi \sqrt{m} \Big/ \sqrt{ \sum_{j=1}^{2} \frac{E_p\{\tilde{w}_k^2 \operatorname{Var}_u(Y_{ijk})\}}{\bar{n}} + \operatorname{Var}_p(\tilde{w}_k E_u(D_k)) },$$

where $\tilde{w}_k \propto N_{1k} + N_{2k}$. Similar to UATE, if population clusters sizes are equal, we obtain a simpler power function

$$
\begin{aligned}
(13) \quad & 1 + \mathcal{T}_{m-1}\left(-t_{m-1,\alpha/2}\,\Big|\, \frac{d_P \sqrt{m}}{\sqrt{1+\pi/\bar{n}}}\right) \\
& - \mathcal{T}_{m-1}\left(t_{m-1,\alpha/2}\,\Big|\, \frac{d_P \sqrt{m}}{\sqrt{1+\pi/\bar{n}}}\right),
\end{aligned}
$$

where, for UATE, $\psi$ is normalized by the standard deviation of the within-pair mean difference, $d_P \equiv \psi/\sqrt{\operatorname{Var}_p\{E_u(D_k)\}}$, and $\pi$ is the ratio of the mean variances of the potential outcomes and the variance of within-pair differences-in-means by the mean variances of the potential outcomes, $\pi \equiv \sum_{j=1}^{2} E_p\{\operatorname{Var}_u \cdot (Y_{ijk})\}/\operatorname{Var}_p(E_u(D_k))$.

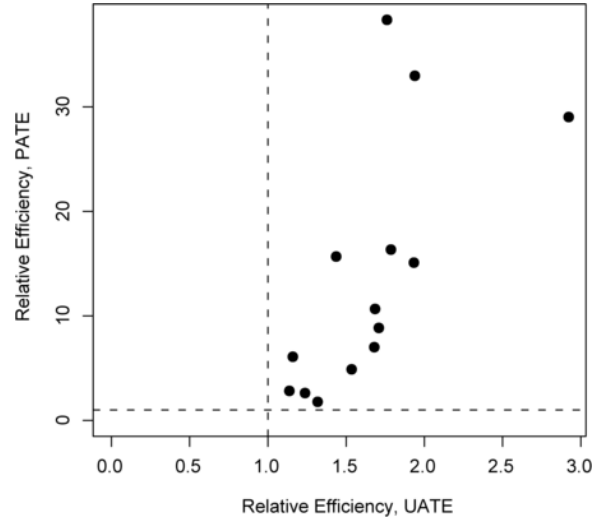

FIG. 2. *Relative efficiency of matched-pair over unmatched cluster randomized designs in the SPS evaluation.*

5.3.2 *Sample size calculations.* We use the above results to estimate the sample size required to achieve a given precision in a future experiment under MPCR. Suppose an investigator wishes to specify the desired degree of precision in terms of Type I and Type II error rates in hypothesis testing, denoted by $\alpha$ and $\beta$, respectively. In particular, the goal is to calculate the sample size required to achieve a given degree of power, $1 - \beta$, against a particular alternative (Snedecor and Cochran, 1989, Section 6.14), using the power functions just derived. For example, for UATE under equal cluster sizes, and using equation (12), the desired number of cluster pairs is the smallest value of $m$ such that $1 + \mathcal{T}_{m-1}(-t_{m-1,\alpha/2} \mid d_U\sqrt{m}) - \mathcal{T}_{m-1}(t_{m-1,\alpha/2} \mid d_U\sqrt{m}) \geq 1-\beta$ where $d_U \equiv \psi/\sqrt{\operatorname{Var}(D_k)}$, $\alpha$, and $\beta$ are specified by the researcher. Similarly, for PATE, equation (13) is used to determine the number of pairs and units within each cluster.

*Empirical evidence.* To illustrate, we use SPS evaluation data on the annualized out-of-pocket health care expenditure that a household spent in the most recent month. Using estimates of $\pi$ and $\operatorname{Var}_p\{E_u(D_k)\}$ from the SPS data and equation (13), we calculate the minimal absolute effect size for PATE that can be detected using a two-sided $t$-test with size 0.95 and power 0.8, for any given cluster size and number of cluster pairs. Since the household is the unit of interest in this example, our population count involves the number of households per cluster, instead of the number of individuals.

In the left panel of Figure 3, horizontal axis is the number of pairs and the vertical axis indicates the number of units within each cluster. The contour lines represent the minimum detectable size in pesos. The graph shows that MPCR with 30 pairs and 100 units within each cluster can detect the true absolute effect size of approximately 450 pesos with the given precision. The figure displays the obvious result that experiments with more pairs or clusters, can detect smaller sized effects (contour lines are labeled with smaller numbers as we move to the top right of the figure). More importantly, the nearly vertical contour lines (above 50 or so units within each cluster) indicates that adding more pairs of clusters adds more statistical power than adding more units within each pair. However, adding one more pair means that many more units will be added, and in some situations sampling units within new clusters is more expensive than within existing clusters. As such, the exact tradeoff depends on the specifics of each application, and it would be incorrect to conclude that more clusters always dominates more units. (We discuss the right panel of the figure next.)

5.3.3 *Power comparison.* Although MPCR is typically more efficient than UMCR regardless of sample size, Martin et al. [[1993], page 330] point out that when the number of pairs is small (fewer than about 10), "the matched design will probably have less power than the unmatched design" due to the loss of degrees of freedom. Here, we show that this conclusion critically hinges on Martin et al.'s assumption of equal cluster population sizes as well as their particular assumed parametric model relating the matching and outcome variables. Modeling assumptions are always worrisome, but the equal cluster size assumption is especially problematic because varying cluster sizes is in fact a fundamental feature of numerous CR experiments.

When cluster sizes are unequal, the efficiency gain of matching in CR trials depends on the correlations of *weighted* cluster means between the treatment and control clusters across pairs (with weights based on sample or population cluster sizes depending on the quantity of interest), not the unweighted correlations used in Martin et al.'s calculations. Since population cluster sizes are typically observed prior to the treatment randomization, researchers can incorporate this variable into their matching procedure. As a result, correlations of weighted outcomes (constructed from clusters with matched weights) will usually be *substantially* higher than those of unweighted outcomes; this is true even when cluster sizes are independent of outcomes. Thus, in CR trials with unequal cluster sizes, the efficiency gain due to pre-randomization matching is likely to be considerably greater than the equal cluster size case considered by Martin et al. (1993). Any power comparison must take this factor into consideration, and along with the bias reduction, this is another reason to incorporate cluster sizes into one's matching procedure.

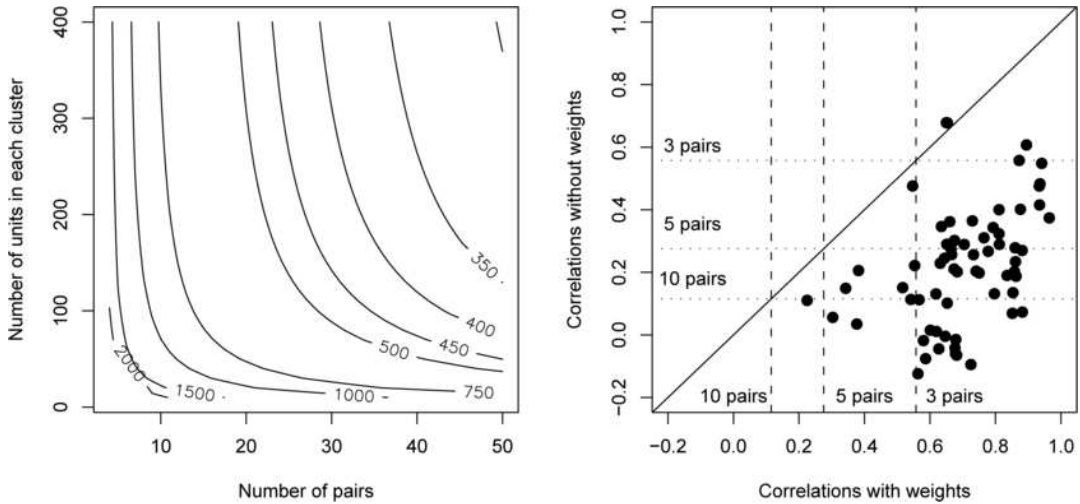

Fig. 3.    *Sample size calculations for PATE under MPCR. The left panel plots the smallest detectable absolute effect size of SPS on annualized out-of-pocket expenditures (in pesos) using a 0.05 level two-sided test with power 0.95, with π estimated from SPS data. The horizontal and vertical axes plot m and n̄, respectively. The right panel compares correlations with and without population weights between treatment and control group cluster-specific means in SPS data. All but one variable has higher correlation when incorporating weights, as seen by a dot below the 45° line. The graph also presents "break-even" correlations (indicated by dashed and dotted lines with and without weights, respectively), which are the smallest possible correlations matching must induce in order for MPCR to detect smaller effect size than the UMCR, given fixed power (0.8) and size (0.95). The graph suggests, when weights are appropriately taken into consideration, that MPCR should be preferred (for all but possibly one variable) even when the number of pairs is as small as three.*

*Empirical evidence.* The right panel of Figure 3 illustrates the argument above using the SPS evaluation data, by calculating the across-pair correlations between treatment and control cluster means of 67 outcome variables (ranging from health related variables to household health care expenditure variables), both with and without weights. We use population cluster sizes as weights, which were observed prior to the randomization of the treatment and incorporated into the matching procedure used (King et al., 2007). The graph shows that all but one variable has considerably higher correlations when weights are incorporated (which does not make the equal cluster size assumption) than when they are ignored (which assumes constant cluster size); this can be seen by all but one of the dots falling below the solid 45° line. In fact, the median of the correlations is more than *three* times larger with (0.68) than without (0.20) weights. In their conclusion, Martin et al. [(1993), page 336] recommend that if the number of pairs is 10 or fewer, then matching should be used only if researchers are confident that the correlation due to matching is at least 0.2. Indeed, all variables in SPS meet this criteria if the weights are appropriately taken into account, the minimum correlation with weights being 0.22. (If the correlations are calculated incorrectly as they did without weights, then only about half of the variables meet their criteria.)

To illustrate the above result in terms of power and sample size calculations, the graph also presents the "break-even" matching correlations (indicated by dashed and dotted lines for correlations with and without weights, respectively) that are used by Martin et al. (1993), Section 7. As in the original article, we set the power and size of the test to be 0.8 and 0.95, respectively, and derive the smallest correlation matching must induce in order for the matched-pair design to be able to detect smaller effect sizes than the UMCR design. The result indicates that even with as few as three pairs, more than 85% of the variables had a correlation higher than the break-even point, which is 0.56. With five pairs, all but one variable exceeds the threshold.

In contrast, if one ignores the weights, by incorrectly assuming that the clusters are equally sized, as in Martin et al., then only 4% and 34% of the variables have the correlations higher than the break-even correlations of three and five pairs, respectively. Martin et al. ([1993](#)) described the correlation of 0.25 as "difficult to achieve by matching" (page 335). However, as the data from SPS evaluation show, since one can match on cluster sizes, the level of *weighted* correlations is much higher when cluster sizes are different.

For another example, Donner and Klar [([2000b](#)), page 37] give the *unweighted* correlations from seven different studies, only one of which is negative (0.49, 0.41, 0.13, 0.63, −0.32, 0.94 and 0.21). The correct weighted correlations are not reported, but in all cases would be higher, and in all likelihood all seven would be positive.

Thus, by dropping the assumption that all clusters are equally sized we have shown here that, for practical purposes, the matched pair design may well have more statistical power than the UMCR design, even for small samples. Of course, if one has fewer than three matched pairs, it's probably time to stop worrying about the properties of statistical estimators and head to the field to gather more data.

### 5.4 Lost Clusters, Stratified Designs and Causal Heterogeneity

We now clarify four additional issues about MPCR that have arisen. First, some recommend a stratified design, where units are matched in blocks of larger than two. However, a stratified design is merely a UMCR design operating within each stratum. If all units within a stratum have identical values on important background covariates, then it is effectively equivalent to MPCR. But if any heterogeneity on these covariates or cluster sizes remain within strata, then the stratified design may leave some efficiency on the table. Thus, when feasible, switching from a stratified to an MPCR design has the potential to greatly increase efficiency and power.

Second, Donner and Klar [([2000a](#)), page 40] explain that a "disadvantage of the matched-pair design is that the loss to follow-up of a single cluster in a pair implies that both clusters in that pair must effectively be discarded from the trial, at least with respect to testing the effect of the intervention. This problem... clearly does not arise if there is some replication of clusters within each combination of intervention and stratum." Indeed, the loss of a single cluster from a stratum with more than two clusters may make it possible to estimate the causal effect within that stratum, but the missingness process must be ascertained or assumed and some type of imputation strategy (or other procedure; e.g., Wei, [1982](#)) must be used, risking model dependence. These are issues for both MPCR and stratified designs. Alternatively, if a cluster is lost in an MPCR study, then dropping the other member of the pair makes it possible to retain the benefits of randomization for SATE or CATE defined in the remaining pairs—without losing other observations, without imputation and possible model dependence, and regardless of the missing data mechanism (King et al., [2007](#)). In contrast, the loss of a cluster in a UMCR design turns an experimental study into an observational study requiring the addition of ignorability assumptions which experimentalists normally try to avoid. The loss of a single cluster within a stratum larger than two units means that more than one cluster will need to be dropped in order to retain the benefits of randomization, which may lead to unnecessary efficiency losses.

Third is the claim that MPCR restricts "prediction models to cluster-level baseline risk factors (for example, cluster size)" (Donner and Klar, [2004](#)). This sentence has been widely interpreted to mean that prediction models under MPCR cannot include baseline risk factors, but Donner and Klar clearly intended it to indicate (and confirmed to us that they meant) the more straightforward point that cluster-level fixed effects cannot be included in regression models under MPCR. Of course, results can be analyzed within strata defined by any individual or cluster level variable, so long as it is pre-treatment. For example, the bottom two rows of Table [3](#) repeat the same analysis as the top two rows but only for male-headed households, a variable measured only at the unit-level and used to separate the sample at that level. (The results for each quantity of interest in this case appear only slightly larger than for the entire sample.) Regression models with fixed effects for clusters are unidentified under MPCR, although substituting in random effects is unproblematic, at least for identification purposes.

Finally, Donner and Klar (2004) explain that MPCR is to be faulted because of its "inability to test for homogeneity" of causal effects within a pair. And hypothesis tests cannot be conducted for the difference between two pairs. However, the causal effect is easy to measure without bias or model dependence under MPCR (but not under UMCR) at the pair level without bias merely by taking the difference in means between the two clusters. This may be a noisy estimate if matching quality is poor, but it serves as a useful unbiased dependent variable for subsequent analyses. We can see how it varies as a function of any variable measured at the unit level and then aggregated to the cluster-pair level, or measured directly at the aggregate level from existing data, such as from census data. Even hypothesis tests are possible if we pool pairs. For example, since the point of SPS was to help poor families, we could examine whether the causal effect of rolling out SPS on various outcome variables increases as the wealth of an area drops. This can be done by a simple plot of the pair-level causal effect by wealth, or fitting a regression model.

## 6. METHODS FOR UNIT-LEVEL NONCOMPLIANCE

CR trials typically have imperfect treatment compliance at the unit level. Some individuals in treatment clusters refuse treatment while others in the control cluster receive the treatment. Since most CR social experiments, including the SPS evaluation, allow noncompliance, analyses, in addition to ITT estimates, may account for noncompliance and estimate the effect of the program only for individuals who would adhere to the experimental protocol. Thus, we now extend our approach to CR trials under the *MPCR encouragement design*, where the encouragement to receive a treatment, rather than the receipt of the treatment itself, is randomized at the cluster-level.

Angrist, Imbens and Rubin (1996) show how an instrumental variable method can be used to analyze unit-randomized experiments with noncompliance under individually randomized designs. We extend their approach to MPCR experiments with unit-level noncompliance. To complement the parametric Bayesian approach to this problem (under the unmatched cluster randomized design) by Frangakis,

Rubin and Zhou (2002), we consider a design-based analysis applying the approach introduced in Section 4.

### 6.1 Causal Quantities of Interest

We consider the two types of causal quantities of interest under MPCR encouragement designs—the intention-to-treat (ITT) effect and the complier average causal effect (CACE) (Angrist, Imbens and Rubin, 1996). The ITT effect is the average causal effect of encouragement (rather than treatment) and is equivalent to the various versions of the average treatment effect in Section 3.3 (i.e., SATE, CATE, UATE and PATE, with or without interference).

In contrast, the CACE estimand is the average treatment effect (for SATE, CATE, UATE or PATE, with or without interference) among compliers only. Compliers are neither those merely observed to affiliate among those in encouragement clusters nor those observed not to affiliate in clusters not encouraged since the former includes always-takers and the latter includes never-takers. Note that always-takers (never-takers) are those who always (never) take the treatment regardless of whether or not they are encouraged. In addition, these groups are defined as a consequence of the treatment. Compliers are those who would affiliate only if they were encouraged and would not affiliate only if they were not encouraged, and so this group is defined prior to the encouragement but its members are not completely observed. We propose a method that can be used to estimate CACE.

### 6.2 Design and Notation

The setup is the same as Section 3.2 except that $T_{jk}$ represents whether the units in the $j$th cluster in the $k$th pair are encouraged to receive the treatment rather than whether it received the treatment. Recall that $T_{1k} = Z_k$ and $T_{2k} = 1 - Z_k$. Now, let $R_{ijk}(T_{jk})$ be the potential treatment receipt indicator variables for the $i$th unit in the $j$th cluster of the $k$th pair under the encouragement ($T_{jk} = 1$) and control ($T_{jk} = 0$) conditions. The observed treatment variable is, then, $R_{ijk} \equiv T_{jk}R_{ijk}(1) + (1 - T_{jk})R_{ijk}(0)$. Similar to the potential outcomes, these potential treatment variables depend on cluster-level encouragement variable rather than the unit-level encouragement variable, requiring a different interpretation of the resulting causal effects. Finally, we write the potential outcomes as functions of (cluster-level) randomized encouragement and actual receipt of treatment (at the unit-level), that is, $Y_{ijk}(R_{ijk}, T_{jk})$. This formulation makes the following assumption, which an extension of Assumption 1:

ASSUMPTION 3 (No interference between units). *Let $R_{ijk}(\mathbf{T})$ be the potential outcomes for the $i$th unit in the $j$th cluster of the $k$th matched-pair where $\mathbf{T}$ is an $(m \times 2)$ matrix whose $(j,k)$ element is $T_{jk}$. Furthermore, let $Y_{ijk}(\mathbf{R}, \mathbf{T})$ be the potential outcomes for the $i$th unit in the $j$th cluster of the $k$th matched-pair where $\mathbf{R}$ is an $(n_{jk} \times m \times 2)$ ragged array whose $(i,j,k)$ element is $R_{ijk}$. Then:*

1. *If $T_{jk} = T'_{jk}$, then $R_{ijk}(\mathbf{T}) = R_{ijk}(\mathbf{T}')$.*
2. *If $T_{jk} = T'_{jk}$ and $R_{ijk} = R'_{ijk}$, then $Y_{ijk}(\mathbf{R}, \mathbf{T}) = Y_{ijk}(\mathbf{R}', \mathbf{T}')$.*

In other words, this assumption requires that one person's decision to affiliate has no effect on any other person's outcomes within the same cluster; as such, the requirements are more demanding than for the ITT effects above. This assumption might be violated for certain health outcomes in the SPS evaluation: if all of one's neighbors affiliate with SPS, the health care they receive may reduce the prevalence of infectious diseases and so might thereby improve that person's health outcomes (an example of "herd immunity"). Relaxing this assumption thus remains an important methodological issue that seems worthy of future research.

The no interference assumption allows us to write $R_{ijk}(\mathbf{T}) = R_{ijk}(T_{jk})$ and $Y_{ijk}(\mathbf{R}, \mathbf{T}) = Y_{ijk}(R_{ijk}, T_{jk})$. Since $T_{1k} = Z_k$ and $T_{2k} = 1 - Z_k$, both $R_{ijk}(T_{jk})$ and $Y_{ijk}(T_{jk})$ depend on $Z_k$ alone.

Extending the framework of Angrist, Imbens and Rubin (1996) to CR trials, we make an exclusion restriction so that cluster-level encouragement affects the unit-level outcome only through the unit-level receipt of the treatment:

ASSUMPTION 4 (Exclusion restriction). *$Y_{ijk}(r, 0) = Y_{ijk}(r, 1)$ for $r = 0, 1$ and all $i, j,$ and $k$.*

These assumptions together simplify the problem by enabling us to write the potential outcomes as functions of $T_{jk}$ (or $Z_k$) alone, that is, $Y_{ijk}(R_{ijk}, T_{jk}) = Y_{ijk}(T_{jk})$.

Finally, following Angrist, Imbens and Rubin (1996), we call the units with $R_{ijk}(T_{jk}) = T_{jk}$ *compliers* (and denote them by $C_{ijk} = c$), those with $R_{ijk}(T_{jk}) = 1$ *always-takers* ($C_{ijk} = a$), those with $R_{ijk}(T_{jk}) = 0$ *never-takers* ($C_{ijk} = n$), and the units with $R_{ijk}(T_{jk}) = 1 - T_{jk}$ *defiers* ($C_{ijk} = d$). The monotonicity assumption excludes the existence of defiers.

ASSUMPTION 5 (Monotonicity). *There exists no defier. That is, $R_{ijk}(1) \geq R_{ijk}(0)$ holds for all $i, j, k$.*

In our Mexico evaluation, never-takers are those who would not affiliate with SPS regardless of whether the government encourages them to do so or not. Since SPS was designed for the poor, many wealthy citizens with their own preexisting health care arrangements may be never-takers. We expected a substantial proportion of the population to qualify as never-takers, and in fact estimate them at 56%. Always-takers are those who would affiliate with SPS regardless of assignment. These are more uncommon, and would likely be the poor without access to health care who nevertheless have the information and financial resources necessary to travel to the place to sign up for SPS and to travel back to receive care. (The estimated proportion of always-takers is only 7%.) The last type is defiers, or people who would affiliate with SPS if not encouraged to do so but would not affiliate if encouraged. Assuming the absence of defiers seems reasonable.

## 6.3 Estimation

If we assume sampling of both pairs of clusters and units within each cluster, then the ITT causal effect can be defined as $\psi_P$. Thus, $\hat{\psi}(N_{1k} + N_{2k})$ can be used to estimate this ITT effect, and the approximately unbiased estimation of its variance is possible using the results given in Section 4.3.

Next, we consider *population* CACE. Under the assumption of simple random sampling of both clusters and units within each cluster, this estimand is defined as $\gamma \equiv \mathbb{E}_{\mathcal{P}}(Y(1) - Y(0)|C = c) = \mathbb{E}_{\mathcal{P}}(Y(1) - Y(0))/\mathbb{E}_{\mathcal{P}}(R(1) - R(0))$, where the equality follows from the direct application Angrist, Imbens and Rubin (1996) to CR trials under the assumptions stated above. If we only assume simple random sampling of clusters as in UATE, then the expectation in $\gamma$ is taken with respect to the set $\mathcal{U}$ rather than $\mathcal{P}$.

Thus, the instrumental variable estimator based on the general weighted estimator in equation (3) is $\hat{\gamma}(w_k) \equiv \hat{\psi}(w_k)/\hat{\tau}(w_k)$, where $\hat{\tau}(w_k)$ is the estimator of the ITT effect on the receipt of the treatment

$$\hat{\tau}(w_k) \equiv \frac{1}{\sum_{k=1}^{m} w_k}$$
$$\cdot \sum_{k=1}^{m} w_k \left\{ Z_k \left( \frac{\sum_{i=1}^{n_{1k}} R_{i1k}}{n_{1k}} - \frac{\sum_{i=1}^{n_{2k}} R_{i2k}}{n_{2k}} \right) \right.$$
$$+ (1 - Z_k)$$
$$\left. \cdot \left( \frac{\sum_{i=1}^{n_{2k}} R_{i2k}}{n_{2k}} - \frac{\sum_{i=1}^{n_{1k}} R_{i1k}}{n_{1k}} \right) \right\}.$$

When matching is effective or when cluster sizes are equal within each matched-pair, this estimator is consistent and approximately unbiased. Using a Taylor series expansion, the variance of this estimator can be approximated by

$$
\begin{aligned}
(14) \quad & \mathrm{Var}_{apu}(\hat{\gamma}(w_k)) \\
& \approx \frac{1}{\{E_{apu}(\hat{\tau}(w_k))\}^4} \\
& \quad \cdot [\{E_{apu}(\hat{\tau}(w_k))\}^2 \, \mathrm{Var}_{apu}(\hat{\psi}(w_k)) \\
& \quad + \{E_{apu}(\hat{\psi}(w_k))\}^2 \, \mathrm{Var}_{apu}(\hat{\tau}(w_k)) \\
& \quad - 2 E_{apu}(\hat{\psi}(w_k)) E_{apu}(\hat{\tau}(w_k)) \\
& \quad \cdot \mathrm{Cov}_{apu}(\hat{\psi}(w_k), \hat{\tau}(w_k))],
\end{aligned}
$$

where if simple random sampling of pairs of clusters alone is assumed, then the subscript "$apu$" (for assignment, pairs, and units) is replaced with "$ap$." Furthermore, the argument given in Section 4.3 implies, for example, that the variance of $\hat{\gamma}(\tilde{w}_k)$ for estimating the *sample* CACE is on average less than the variance for the *population* CACE given in equation (14).

Finally, Proposition 3 shows how to estimate $\mathrm{Var}_{apu}(\hat{\psi}(w_k))$, $\mathrm{Var}_{apu}(\hat{\tau}(w_k))$ (or $\mathrm{Var}_{ap}(\hat{\psi}(w_k))$ and $\mathrm{Var}_{ap}(\hat{\tau}(w_k))$) approximately without bias. Thus, we only need an estimate of the covariance between $\hat{\psi}(w_k)$ and $\hat{\tau}(w_k)$ from the observed data. Using the normalized weights $\tilde{w}_k$, Appendix A.5 proves that the following estimator is approximately unbiased for both $\mathrm{Cov}_{apu}(\hat{\psi}(w_k), \hat{\tau}(w_k))$ and $\mathrm{Cov}_{pu}(\hat{\psi}(w_k), \hat{\tau}(w_k))$ under their respective sampling assumptions:

$$
\begin{aligned}
& \hat{\nu}(\tilde{w}_k) \\
& \equiv \frac{m}{(m-1)n^2} \\
& \quad \cdot \sum_{k=1}^{m} \Bigg[ \tilde{w}_k \Bigg\{ Z_k \Bigg( \frac{\sum_{i=1}^{n_{1k}} Y_{i1k}}{n_{1k}} - \frac{\sum_{i=1}^{n_{2k}} Y_{i2k}}{n_{2k}} \Bigg) \\
& \qquad + (1 - Z_k) \\
& \qquad \cdot \Bigg( \frac{\sum_{i=1}^{n_{2k}} Y_{i2k}}{n_{2k}} - \frac{\sum_{i=1}^{n_{1k}} Y_{i1k}}{n_{1k}} \Bigg) \Bigg\} \\
& \qquad - \frac{n\hat{\psi}(\tilde{w}_k)}{m} \Bigg] \\
& \quad \cdot \Bigg[ \tilde{w}_k \Bigg\{ Z_k \Bigg( \frac{\sum_{i=1}^{n_{1k}} R_{i1k}}{n_{1k}} - \frac{\sum_{i=1}^{n_{2k}} R_{i2k}}{n_{2k}} \Bigg) \\
& \qquad + (1 - Z_k) \Bigg( \frac{\sum_{i=1}^{n_{2k}} R_{i2k}}{n_{2k}} - \frac{\sum_{i=1}^{n_{1k}} R_{i1k}}{n_{1k}} \Bigg) \Bigg\}
\end{aligned}
$$

$$
\qquad - \frac{n\hat{\tau}(\tilde{w}_k)}{m} \Bigg].
$$

## 7. SEGURO POPULAR EVALUATION

We now estimate the causal effect of SPS on the probability of a household suffering catastrophic health expenditures (out-of-pocket health care expenditures totaling more than 30% of a household's annual post-subsistence or disposable income). As nearly 10% of households suffer catastrophic health expenditures in a year, it is easy to see why this would be a major priority. We estimate all four target population quantities of interest (SATE, CATE, UATE, and PATE) both for the intention to treat (ITT) effect of encouragement to affiliate an the average causal effect among compliers (CACE). Although in most applications, substantive interest would narrow this list to one or a few of these quantities, for our methodological purposes we present all eight estimates (and standard errors) in Table 3.

A table like this will always have some of the same features, no matter what variable is analyzed. Recall, for example, that point estimates of SATE and UATE are the same, as they are for CATE and PATE. In addition, standard errors of UATE and PATE are the upper bounds of the standard errors for SATE and CATE, respectively. CACE estimates of course are never smaller than those for ITT.

For the specific estimates, consider first the two top lines of Table 3 corresponding to all households. For these data, the CACE estimates are about 2.7 times larger than that for ITT. The large difference is because of all those who had preexisting health care and so were largely never-takers. Overall, these results indicate that SPS was clearly successful in reducing the most devastating type of medical expenditures. The differences among the columns indicate that the average causal effect of encouragement to affiliate to SPS (the ITT effect) is somewhat larger in the population of individuals represented by our sample ($-0.023$) than among the individuals we directly observe ($-0.014$). The same is also true among compliers, but at a higher level ($-0.038$ vs. $-0.064$).

Substantively, these numbers are quite large. Since those who suffer from catastrophic health expenditures are mostly the poor without access to health insurance, they are likely to be disproportionately represented among compliers as compared to the wealthy with preexisting health care arrangements.

Table 3
*Estimates of eight causal effect of SPS on the probability of catastrophic health expenditures for all households and male-headed households (standard errors in parentheses)*

|             |      | SATE | CATE | UATE | PATE |
|-------------|------|------|------|------|------|
| All         | ITT  | $-0.014$ ($\leq 0.007$) | $-0.023$ ($\leq 0.015$) | $-0.014$ (0.007) | $-0.023$ (0.015) |
|             | CACE | $-0.038$ ($\leq 0.018$) | $-0.064$ ($\leq 0.024$) | $-0.038$ (0.018) | $-0.064$ (0.024) |
| Male-headed | ITT  | $-0.016$ ($\leq 0.008$) | $-0.025$ ($\leq 0.018$) | $-0.016$ (0.008) | $-0.025$ (0.018) |
|             | CACE | $-0.042$ ($\leq 0.020$) | $-0.070$ ($\leq 0.031$) | $-0.042$ (0.020) | $-0.070$ (0.031) |

As such, this analysis indicates that the causal effect of rolling out the policy reduces by about 23% the proportion of those who experience catastrophic expenditures (i.e., $-0.023$ of the 10% with catastrophic expenditures). (Detailed analyses of these and other substantive results from the SPS evaluation appear in King et al., 2009.)

## 8. CONCLUDING REMARKS

The methods developed here are designed for researchers lucky enough to be able to randomize treatment assignment, but stuck because of political or other constraints with having to randomize clusters of individuals rather than the individuals themselves. Field experiments in particular frequently require cluster randomization. Individual-level randomization was impossible in our evaluation of the Mexican SPS program; in fact, negotiations with the Mexican government began with the presumption that no type of randomization would be politically feasible, but it eventually concluded by allowing cluster-level randomization to be implemented.

When clusters of individuals are randomized rather than the individuals themselves, the best practice should involve three steps. First, researchers should choose their causal quantity of interest, as defined in Section 3.3. They should then identify available pre-treatment covariates likely to affect the outcome variable, and, if possible, pair clusters based on the similarity of these covariates and cluster sizes; this step is severely underutilized and, when feasible, will translate into considerable research resources saved and numerous observations gained. Finally, researchers should randomly choose one treated and one control cluster within each pair. Claims in the literature about problems with matched-pair cluster randomization designs are misguided: clusters should be paired prior to randomization when considered from the perspective of efficiency, power, bias or robustness.

Of course, administrative, political, ethical and other issues will sometimes constrain the ability of researchers to pair clusters prior to randomization. With the results and new estimators offered here, the effort in the design of cluster-randomized experiments can now shift from debates about when pairing is useful to practical discussions of how best to marshal creative arguments and procedures to ensure that clusters can more often be paired prior to randomization.

Cornfield [(1978), pages 101–102] concludes his now classic study by writing that "Randomization by cluster accompanied by an analysis appropriate to randomization by individual is an exercise in self-deception, ... and should be avoided," and an enormous literature has grown in many fields echoing this warning. We can now add that randomization by cluster without prior construction of matched pairs, when pairing is feasible, is an exercise in self-destruction. Failing to match can greatly reduce efficiency, power and robustness, and is equivalent to discarding a large portion of experimental data or wasting grant money and investigator effort. This result should affect practice, especially in literatures like political science where experimental analyses routinely use cluster-randomization but examples of matched-pair designs have almost never been used, as well as community consensus recommendations for best practices in the conduct and analysis of cluster-randomized experiments, which closely follow prior methodological literature. These include the extension to the "CONSORT" agreement among the major biomedical journals (Campbell, Elbourne and Altman, 2004), the Cochrane Collaboration requirements for reviewing research (Higgins and Green, 2006, Section 8.11.2), the prominent Medical Research Council (2002) guidelines, and the education research What Works Clearinghouse (2006). Each would seem to require crucial modifications in light of the results given here.

# APPENDIX A: MATHEMATICAL APPENDIX

## A.1 Proof of Proposition 1

This proof uses a strategy similar to that of Proposition 1 of Imai (2008). First, rewrite $\hat{\sigma}(\tilde{w}_k)$ as

$$\frac{(m-1)n^2}{m}\hat{\sigma}(\tilde{w}_k)$$

$$= \sum_{k=1}^{m}\left[\tilde{w}_k\{Z_kD_k(1) + (1-Z_k)D_k(0)\}\right.$$

$$-\frac{1}{m}\sum_{k'=1}^{m}\tilde{w}_{k'}\{Z_{k'}D_{k'}(1)$$

$$\left. + (1-Z_{k'})D_{k'}(0)\}\right]^2$$

$$= \frac{m-1}{m}\sum_{k=1}^{m}\tilde{w}_k^2\{Z_kD_k(1)^2 + (1-Z_k)D_k(0)^2\}$$

$$-\frac{1}{m}\sum_{k=1}^{m}\sum_{k'\neq k}\tilde{w}_k\tilde{w}_{k'}\{Z_kZ_{k'}D_k(1)D_{k'}(1)$$

$$+ Z_k(1-Z_{k'})D_k(1)D_{k'}(0)$$

$$+ (1-Z_k)Z_{k'}D_k(0)D_{k'}(1)$$

$$+ (1-Z_k)(1-Z_{k'})D_k(0)$$

$$\cdot D_{k'}(0)\}.$$

Assumption 2 implies $E_a(Z_k) = 1/2$ and $E_a(Z_kZ_{k'}) = 1/4$ for $k \neq k'$. Thus, taking expectations over $Z_k$ and rearranging, gives

$$E_a(\hat{\sigma}(\tilde{w}_k))$$

$$(15) \qquad = \frac{1}{2n^2}\left\{\sum_{k=1}^{m}\tilde{w}_k^2(D_k(1)^2 + D_k(0)^2)\right.$$

$$-\frac{1}{2(m-1)}\sum_{k=1}^{m}\sum_{k'\neq k}\tilde{w}_k\tilde{w}_{k'}$$

$$\cdot (D_k(1) + D_k(0))$$

$$\left. \cdot (D_{k'}(1) + D_{k'}(0))\right\}.$$

Finally, we compare this with the true variance expression in (7): $E_a(\hat{\sigma}(\tilde{w}_k)) - \text{Var}_a(\hat{\psi}(\tilde{w}_k))$, which equals

$$\frac{1}{4n^2}\left\{\sum_{k=1}^{m}\tilde{w}_k^2\{D_k(1) + D_k(0)\}^2\right.$$

$$-\frac{1}{m-1}\sum_{k=1}^{m}\sum_{k'\neq k}\tilde{w}_k\tilde{w}_{k'}(D_k(1) + D_k(0))$$

$$\cdot (D_{k'}(1) + D_{k'}(0))\bigg\}$$

$$= \frac{m}{4n^2}\text{var}\{\tilde{w}_k(D_k(1) + D_k(0))\}.$$

This bias term is not identifiable because $D_k(1)$ and $D_k(0)$ are not jointly observed for any $k$, implying that the variance is not identifiable either.

## A.2 Proof of Proposition 2

Applying the law of iterated expectations to equation (15), we have

$$E_{au}(\hat{\sigma}(\tilde{w}_k))$$

$$= \frac{1}{2n^2}\left[\sum_{k=1}^{m}\tilde{w}_k^2E_u\{D_k(1)^2 + D_k(0)^2\}\right.$$

$$(16) \qquad -\frac{1}{2(m-1)}$$

$$\cdot \sum_{k=1}^{m}\sum_{k'\neq k}\tilde{w}_k\tilde{w}_{k'}E_u(D_k(1) + D_k(0))$$

$$\left. \cdot E_u(D_{k'}(1) + D_{k'}(0))\right],$$

where the equality follows from the assumption that sampling of units is independent across clusters. Together with the definition of $\text{Var}_{au}(\hat{\psi}(\tilde{w}_k))$ given above, we have

$$E_{au}(\hat{\sigma}(\tilde{w}_k)) - \text{Var}_{au}(\hat{\psi}(\tilde{w}_k))$$

$$= \frac{1}{4n^2}\left[\sum_{k=1}^{m}\tilde{w}_k^2\{E_u(D_k(1)^2 + D_k(0)^2)\right.$$

$$- \text{Var}_u(D_k(1)) - \text{Var}_u(D_k(0))$$

$$+ 2E_u(D_k(1))E_u(D_k(0))\}$$

$$-\frac{1}{m-1}$$

$$\cdot \sum_{k=1}^{m}\sum_{k'\neq k}\tilde{w}_k\tilde{w}_{k'}E_u(D_k(1) + D_k(0))$$

$$\left. \cdot E_u(D_{k'}(1) + D_{k'}(0))\right]$$

$$= \frac{m}{4n^2}\text{var}\{\tilde{w}_kE_u(D_k(1) + D_k(0))\}.$$

Since we do not observe $D_k(1)$ and $D_k(0)$ jointly, this variance is not identified.

### A.3 Proof of Proposition 3

Since UATE is a special case of PATE where all units within each cluster are observed ($n_{jk} = N_{jk}$), we first derive the variance of $\hat{\psi}(\tilde{w}_k)$ for PATE. Let $\tilde{D}_k(t) = w_k D_k(t)$, $\tilde{\mu}_k(t) = E_u(\tilde{D}_k(t))$, and $\tilde{\eta}_k(t) = \text{Var}_u(\tilde{D}_k(t))$ for $t = 0, 1$. Then, randomizing the order of clusters within each pair implies $E_c(\tilde{\eta}_k) = E_c(\tilde{\eta}_k(1)) = E_c(\tilde{\eta}_k(0))$ and $\text{Var}_c(\tilde{\mu}_k) = \text{Var}_c(\tilde{\mu}_k(1)) = \text{Var}_c(\tilde{\mu}_k(0))$. Then, the variance is

$$\text{Var}_{apu}(\hat{\psi}(\tilde{w}_k))$$
$$= \frac{1}{2m\bar{w}^2} E_p \Big[ \text{Var}_u(\tilde{D}_k(1)) + \text{Var}_u(\tilde{D}_k(0))$$
$$+ \frac{1}{2} \{ E_u(\tilde{D}_k(1) - \tilde{D}_k(0)) \}^2 \Big]$$
$$+ \frac{1}{4m\bar{w}^2} \text{Var}_p(E_u(\tilde{D}_k(1) + \tilde{D}_k(0)))$$
$$= \frac{1}{m\bar{w}^2} \{ E_p(\tilde{\eta}_k) + \text{Var}_p(\tilde{\mu}_k) \}.$$

When the estimand is UATE, $\tilde{\eta}_k = 0$ for all $k$ since within-cluster means are observed without sampling variability. Thus, $\text{Var}_{apu}(\hat{\psi}(\tilde{w}_k)) = \text{Var}_p(\tilde{\mu}_k)/(m\bar{w}^2)$.

Next, we show that $\hat{\sigma}(\tilde{w}_k)$ is approximately unbiased by applying the law of iterated expectations to Equation (16):

$$E_{apu}(\hat{\sigma}(\tilde{w}_k))$$
$$= \frac{1}{2m\bar{w}^2} \Big[ E_p\{ w_k^2 E_u(D_k(1)^2 + D_k(0)^2) \}$$
$$- \frac{1}{2} [E_p\{ w_k E_u(D_k(1) + D_k(0)) \}]^2 \Big]$$
$$= \frac{1}{m\bar{w}^2} [E_p\{ \text{Var}_u(\tilde{D}_k) \} + \text{Var}_p(\tilde{\mu}_k)],$$

where $E_u(D_k^2) = E_u(D_k(0)^2) = E_u(D_k(1)^2)$ holds because the order of clusters within each pair is randomized. For PATE, $\text{Var}_u(D_k) = 0$ for all $k$ since within-cluster means are observed without sampling uncertainty. Thus, $E_{ap}(\hat{\sigma}(\tilde{w}_k)) = \text{Var}_p(\tilde{\mu}_k)/(m\bar{w}^2)$.

### A.4 Properties of the Harmonic Mean Estimator and Standard Error

*Modeling assumptions.* The harmonic mean estimator, with weights based on the harmonic mean of sample cluster sizes $w_k = n_{1k}n_{2k}/(n_{1k} + n_{2k})$, stems from the weighted one-sample $t$-test for the difference in means: $D_k \overset{\text{indep.}}{\sim} \mathcal{N}(\mu, (w_k/\sum_{k'=1}^m w_{k'})^{-1}\sigma)$ for $k = 1, 2, \ldots, m$ where $w_k$ is the known harmonic mean weight. In our context, $D_k$ is the observed within-pair mean difference, that is, $D_k \equiv Z_k D_k(1) + (1 - Z_k)D_k(0)$ where $D_k(1) \equiv \sum_{i=1}^{n_{1k}} Y_{i1k}(1)/n_{1k} - \sum_{i=1}^{n_{2k}} Y_{i2k}(0)/n_{2k}$ and $D_k(0) \equiv \sum_{i=1}^{n_{2k}} Y_{i2k}(0)/n_{2k} - \sum_{i=1}^{n_{1k}} Y_{i1k}(0)/n_{1k}$. It is well known that under this model, $\sum_{k=1}^m w_k D_k / \sum_{k'=1}^m w_{k'}$ is the uniformly minimum variance unbiased estimator.

Although the derivation of this model is not discussed in the cluster randomization literature, a model commonly used in the statistics literature for other purposes gives rise to these weights (see e.g., Kalton, 1968): $Y_{ijk}(t) \overset{\text{i.i.d.}}{\sim} \mathcal{N}(\mu_t, \tilde{\sigma})$ for $t = 0, 1$ where $\tilde{\sigma} = \sigma \sum_{k=1}^m w_k$ and $\sum_{k=1}^m w_k$ is a known constant since $w_k$ is assumed fixed. The normality assumption is not necessary for some inferential purposes, but this model does require (1) independent and identical distributions across units within each cluster as well as (2) across clusters and pairs (which of course implies constant means and variances within and across clusters and pairs) and (3) equal variances for the two potential outcomes. In sum, the model assumes homogeneity within and across matched-pairs. [Although we focus on the $t$-test here, for binary outcomes the suggested approach in the literature is also based on a homogeneity assumption where the odds ratio is assumed constant across clusters; see, e.g., Donner and Donald (1987); Donner and Hauck, (1989).]

*Bias conditions.* The harmonic mean weight differs from our proposed weight in three ways. First, it gives more weight to pairs with well-matched cluster sizes than to pairs whose cluster sizes are unbalanced. That is, if we assume the sum $n_{1k} + n_{2k}$ is fixed, the harmonic mean is the largest when $n_{1k} = n_{2k}$ and becomes smaller as $n_{1k} - n_{2k}$ increases. Second, and most importantly, this weight does not remove the bias when within-cluster average treatment effects are identical within pairs (so long as heterogeneity across matched-pairs remains), meaning that bias may remain even when matching is effective. (The direction of the bias depends on the data.) One condition under which it is unbiased is with exact matching on sample cluster sizes (i.e., $n_{1k} = n_{2k}$ for all $k$), in which case this estimator coincides with our proposed estimator. Finally, since the weight is based on sample cluster sizes, this estimator is not valid for estimating CATE or PATE.

When its assumptions hold, the harmonic mean estimator is uniformly minimum variance unbiased, and is clearly useful in those circumstances.

*Bias in the variance estimator.* We show here that the variance estimator proposed in the literature (see, e.g., Donner, 1987; Donner and Donald, 1987; Donner and Klar, 1993) may be biased regardless of choice of weights and the direction of bias is indeterminate. A condition under which this variance estimator is unbiased (and approximately equal to ours) is when $m$ is large and the weights are identical across pairs, which is uncommon in practice. We first write this estimator using our notation:

$$
\begin{aligned}
\hat{\delta}(\tilde{w}_k) &\equiv \frac{\sum_{k=1}^{m} \tilde{w}_k^2}{n^3} \\
&\cdot \sum_{k=1}^{m} \tilde{w}_k \Bigg\{ Z_k \left( \frac{\sum_{i=1}^{n_{1k}} Y_{i1k}}{n_{1k}} - \frac{\sum_{i=1}^{n_{2k}} Y_{i2k}}{n_{2k}} \right) \\
&\qquad + (1 - Z_k) \\
&\qquad \cdot \left( \frac{\sum_{i=1}^{n_{2k}} Y_{i2k}}{n_{2k}} - \frac{\sum_{i=1}^{n_{1k}} Y_{i1k}}{n_{1k}} \right) \\
&\qquad - \hat{\psi}(\tilde{w}_k) \Bigg\}^2 .
\end{aligned}
$$
(17)

Next, we rewrite $\hat{\delta}(\tilde{w}_k)$ as

$$
\begin{aligned}
&\frac{n^3}{\sum_{k=1}^{m} \tilde{w}_k^2} \hat{\delta}(\tilde{w}_k) \\
&= \sum_{k=1}^{m} \tilde{w}_k \Bigg[ Z_k D_k(1) + (1 - Z_k) D_k(0) \\
&\qquad - \frac{1}{n} \sum_{k'=1}^{m} \tilde{w}_{k'} \{ Z_{k'} D_{k'}(1) \\
&\qquad\qquad\qquad + (1 - Z_{k'}) D_{k'}(0) \} \Bigg]^2 \\
&= \sum_{k=1}^{m} \tilde{w}_k \Bigg[ Z_k D_k(1)^2 + (1 - Z_k) D_k(0)^2 \\
&\qquad - \frac{2}{n} \sum_{k'=1}^{m} \tilde{w}_{k'} \{ Z_k D_k(1) \\
&\qquad\qquad\qquad + (1 - Z_k) D_k(0) \} \\
&\qquad\qquad \cdot \{ Z_{k'} D_{k'}(1) \\
&\qquad\qquad\qquad + (1 - Z_{k'}) D_{k'}(0) \} \\
&\qquad + \frac{1}{n^2} \sum_{k'=1}^{m} \sum_{k''=1}^{m} \tilde{w}_{k'}^2 \tilde{w}_{k''}^2
\end{aligned}
$$

$$
\begin{aligned}
&\qquad\qquad \cdot \{ Z_{k'} D_{k'}(1) \\
&\qquad\qquad\qquad + (1 - Z_{k'}) D_{k'}(0) \} \\
&\qquad\qquad \cdot \{ Z_{k''} D_{k''}(1) \\
&\qquad\qquad\qquad + (1 - Z_{k''}) D_{k''}(0) \} \Bigg] .
\end{aligned}
$$

Taking the expectation with respect to $Z_k$, $E_a(\hat{\delta}(\tilde{w}_k))$, gives

$$
\begin{aligned}
&\frac{\sum_{k=1}^{m} \tilde{w}_k^2}{2n^3} \sum_{k=1}^{m} \Bigg\{ \left( 1 - \frac{\tilde{w}_k}{n} \right) \tilde{w}_k (D_k(1)^2 + D_k(0)^2) \\
&\qquad - \frac{1}{2n} \sum_{k=1}^{m} \sum_{k' \neq k} \tilde{w}_k \tilde{w}_{k'} (D_k(1) + D_k(0)) \\
&\qquad\qquad\qquad \cdot (D_{k'}(1) + D_{k'}(0)) \Bigg\} .
\end{aligned}
$$

Comparing this expression with $E_a(\hat{\sigma}(\tilde{w}_k))$ in equation (15) shows a difference which remains even after taking the expectation with respect to simple random sampling of pairs of clusters or units within clusters. Since $\hat{\sigma}(\tilde{w}_k)$ is an approximately unbiased estimate of the variance for UATE and PATE, $\hat{\delta}(\tilde{w}_k)$ may be biased.

### A.5 Covariance Estimation

This Appendix derives unbiased estimates of $\mathrm{Cov}_{auc}(\hat{\psi}(\tilde{w}_k), \hat{\tau}(\tilde{w}_k))$ and $\mathrm{Cov}_{ac}(\hat{\psi}(\tilde{w}_k), \hat{\tau}(\tilde{w}_k))$ using the proofs in Propositions 1–3. First, we derive the true covariance between $\hat{\psi}(\tilde{w}_k)$ and $\hat{\tau}(\tilde{w}_k)$. Define $G_k(1) = \sum_{i=1}^{n_{1k}} R_{i1k}(1)/n_{1k} - \sum_{i=1}^{n_{2k}} R_{i2k}(0)/n_{2k}$ and $G_k(0) = \sum_{i=1}^{n_{2k}} R_{i2k}(1)/n_{2k} - \sum_{i=1}^{n_{1k}} R_{i1k}(0)/n_{1k}$. Taking the expectation of with respect to $Z_k$ yields: $\mathrm{Cov}_a(\hat{\psi}(\tilde{w}_k), \hat{\tau}(\tilde{w}_k)) = \frac{1}{n^2} \sum_{k=1}^{m} \tilde{w}_k^2 (D_k(1) - D_k(0)) \cdot (G_k(1) - G_k(0))$. Then, we have

$$
\begin{aligned}
&\mathrm{Cov}_{ap}(\hat{\psi}(\tilde{w}_k), \hat{\tau}(\tilde{w}_k)) \\
&= E_p \{ \mathrm{Cov}_a(\hat{\psi}(\tilde{w}_k), \hat{\tau}(\tilde{w}_k)) \} \\
&\qquad + \mathrm{Cov}_p \{ E_a(\hat{\psi}(\tilde{w}_k)), E_a(\hat{\tau}(\tilde{w}_k)) \} \\
&= \frac{1}{m \bar{w}^2} \mathrm{Cov}_p(\tilde{D}_k, \tilde{G}_k),
\end{aligned}
$$

where $\tilde{G}_k(t) = w_k G_k(t)$ for $t = 0, 1$, and the last equality follows from the fact that $E_p(\tilde{D}_k) = E_p(\tilde{D}_k(t))$, $E_p(\tilde{G}_k) = E_p(\tilde{G}_k(t))$ and $E_p(\tilde{D}_k \tilde{G}_k) = E_p(\tilde{D}_k(t) \tilde{G}_k(t))$ for $t = 0, 1$. Similarly,

$$
\mathrm{Cov}_{au}(\hat{\psi}(\tilde{w}_k), \hat{\tau}(\tilde{w}_k))
$$

$$= E_u\{\mathrm{Cov}_a(\hat{\psi}(\tilde{w}_k), \hat{\tau}(\tilde{w}_k))\}$$
$$+ \mathrm{Cov}_u\{E_a(\hat{\psi}(\tilde{w}_k)), E_a(\hat{\tau}(\tilde{w}_k))\}$$
$$= \frac{1}{2\bar{w}^2}\sum_{k=1}^{m}\Big[\mathrm{Cov}_u(\tilde{D}_k(1), \tilde{G}(1))$$
$$+ \mathrm{Cov}_u(\tilde{D}_k(0), \tilde{G}_k(0))$$
$$+ \frac{1}{2}\{E_u(\tilde{D}_k(1)) - E_u(\tilde{D}_k(0))\}$$
$$\cdot \{E_u(\tilde{G}_k(1)) - E_u(\tilde{G}_k(0))\}\Big].$$

And thus,

$$\mathrm{Cov}_{apu}(\hat{\psi}(\tilde{w}_k), \hat{\tau}(\tilde{w}_k))$$
$$= \frac{1}{m\bar{w}^2}\Big[\mathrm{Cov}_u(\tilde{D}_k, \tilde{G}_k)$$
$$+ \frac{1}{4}E_p\{E_u(\tilde{D}_k(1)) - E_u(\tilde{D}_k(0))\}$$
$$\cdot \{E_u(\tilde{G}_k(1)) - E_u(\tilde{G}_k(0))\}$$
$$+ \frac{1}{4}\mathrm{Cov}_p\{E_u(\tilde{D}_k(1) + \tilde{D}_k(0)),$$
$$E_u(\tilde{G}_k(1) + \tilde{G}_k(0))\}\Big]$$
$$= \frac{1}{m\bar{w}^2}\Big[E_p\{\mathrm{Cov}_u(\tilde{D}_k, \tilde{G}_k)\}$$
$$+ \mathrm{Cov}_p\{E_u(\tilde{D}_k), E_u(\tilde{G}_k)\}\Big].$$

Then, calculations analogous to the ones above shows that $E_{ap}(\hat{\nu}(\tilde{w}_k)) = \mathrm{Cov}_{ap}(\hat{\psi}(\tilde{w}_k), \hat{\tau}(\tilde{w}_k))$ and $E_{apu}(\hat{\nu}(\tilde{w}_k)) = \mathrm{Cov}_{apu}(\hat{\psi}(\tilde{w}_k), \hat{\tau}(\tilde{w}_k))$.

## ACKNOWLEDGMENTS

Our proposed methods can be implemented using an R package, *experiment*, which is available for download at the Comprehensive R Archive Network (http://cran.r-project.org). We thank Kevin Arceneaux, Jake Bowers, Paula Diehr, Ben Hansen, Jennifer Hill, and Dylan Small for helpful comments, and Neil Klar and Allan Donner for detailed suggestions and gracious extended conversations. For research support, our thanks go to the National Institute of Public Health of Mexico, the Mexican Ministry of Health, the National Science Foundation (SES-0550873, SES-0752050), the Princeton University Committee on Research in the Humanities and Social Sciences and the Institute for Quantitative Social Science at Harvard.